\documentstyle[12pt,epsf]{article}
\def\PRL #1 #2 #3 {Phys.\ Rev.\ Lett.\ {\bf #1}, #2 (#3)}
\def\PRD #1 #2 #3 {Phys.\ Rev.\ D~{\bf #1}, #2 (#3)}
\def\PLB #1 #2 #3 {Phys.\ Lett.\ B~{\bf #1}, #2 (#3)}
\def\NPB #1 #2 #3 {Nucl.\ Phys.\ {\bf B#1}, #2 (#3)}
\def\ZPC #1 #2 #3 {Z.\ Phys.\ C~{\bf #1}, #2 (#3)}

\newcommand{\gtap}{\;{\raise.3ex\hbox{$>$\kern-.75em\lower1ex\hbox{$\sim$}}}\;}
\newcommand{\ltap}{\;{\raise.3ex\hbox{$<$\kern-.75em\lower1ex\hbox{$\sim$}}}\;}

\begin{document}
\begin{titlepage}

\rightline{hep-ph/0106281}
\smallskip
\bigskip\bigskip
\begin{center}
{\Large \bf Scale of fermion mass generation} \\
\medskip
\bigskip\bigskip\bigskip
{{\bf F.~Maltoni}, {\bf J.~M.~Niczyporuk}, and {\bf S.~Willenbrock}} \\
\medskip
Department of Physics \\
University of Illinois at Urbana-Champaign \\ 1110 West Green Street \\
Urbana, IL\ \ 61801 \\
\bigskip
\end{center}
\bigskip\bigskip\bigskip

\begin{abstract}
Unitarity of longitudinal weak vector boson scattering implies an upper bound
on the scale of electroweak symmetry breaking, $\Lambda_{EWSB}\equiv
\sqrt{8\pi}v\approx$ 1 TeV.  Appelquist and Chanowitz have derived an
analogous upper bound on the scale of fermion mass generation, proportional to
$v^2/m_f$, by considering the scattering of same-helicity fermions into pairs
of longitudinal weak vector bosons in a theory without a standard Higgs boson.
We show that there is no upper bound, beyond that on the scale of electroweak
symmetry breaking, in such a theory.  This result is obtained by considering
the same process, but with a large number of longitudinal weak vector bosons
in the final state. We further argue that there is no scale of (Dirac) fermion
mass generation in the standard model. In contrast, there is an upper bound on
the scale of Majorana-neutrino mass generation, given by $\Lambda_{Maj}\equiv
4\pi v^2/m_\nu$.  In general, the upper bound on the scale of fermion mass
generation depends on the dimensionality of the interaction responsible for
generating the fermion mass.  We explore the scale of fermion mass generation
in a variety of excursions from the standard model: models with fermions in
nonstandard representations, a theory with higher-dimension interactions, a
two-Higgs-doublet model, and models without a Higgs boson.
\end{abstract}
\end{titlepage}

\newpage

\section{Introduction}

One of the main aspirations of particle physics this decade is the elucidation
of the mechanism that breaks the electroweak gauge symmetry, $SU(2)_L\times
U(1)_Y$, down to the gauge symmetry of electromagnetism, $U(1)_{EM}$.  An
upper bound on the scale of electroweak symmetry breaking,
$\Lambda_{EWSB}\approx 1$ TeV, ensures that the physics of this mechanism is
within reach of the CERN Large Hadron Collider, and perhaps also the Fermilab
Tevatron (if some or all of this physics is much lighter than 1 TeV).
Additional high-energy colliders, such as an $e^+e^-$ linear collider or a
$\mu^+\mu^-$ collider, may be required to completely elucidate the mechanism
of electroweak symmetry breaking.

The upper bound on the scale of electroweak symmetry breaking may be obtained
by considering elastic scattering of longitudinal weak vector bosons.  In the
absence of an explicit model of electroweak symmetry breaking, this amplitude
grows quadratically with energy and violates unitarity at an energy
$\Lambda_{EWSB} \equiv \sqrt{8\pi} v \approx 1$ TeV, where $v= (\sqrt 2
G_F)^{-1/2} \approx 246$ GeV \cite{Chanowitz:1985hj}.  One interprets this as
the scale before which the effective field theory of massive weak vector
bosons must be subsumed by a deeper theory that contains a mechanism for
electroweak symmetry breaking, thereby generating the masses of the weak
bosons.

Appelquist and Chanowitz observed that a similar argument can be put forward
for the scale of fermion mass generation \cite{Appelquist:1987cf}. The
amplitude for scattering of a fermion-antifermion pair of same helicity into a
pair of longitudinal weak vector bosons, in the absence of an explicit model
of fermion mass generation, is proportional to $m_f\sqrt s/v^2$, where $m_f$ is
the fermion mass and $\sqrt s$ is the center-of-mass energy.  This amplitude
violates unitarity at the scale $\Lambda_f \approx v^2/m_f$, which varies with
each fermion depending on its mass and is greater than $\Lambda_{EWSB}$ for
all known fermions. This scale was interpreted as an upper bound on the scale
of fermion mass generation.

Appelquist and Chanowitz noted that there is no known model of fermion mass
generation that saturates the upper bound set by $\Lambda_f$.  This issue was
revisited by Golden, with a similar conclusion \cite{Golden:1994pj}. Attempts
to saturate this bound by considering a two-Higgs-doublet model were also
unsuccessful \cite{Jager:1998va,Chivukula:1998zn}.  However, we recently
showed that a similar upper bound on the scale of Majorana-neutrino mass
generation, $\Lambda_{Maj} \approx v^2/m_\nu$, can be naturally saturated in
explicit models \cite{Maltoni:2001iq}. Given this set of results, one is led
to ask whether the scale $\Lambda_f$ is truly relevant for ordinary (Dirac)
fermions.

In this paper we explore the scale of fermion mass generation in depth.  We
clarify the interpretation of the scale $\Lambda_f$, and we show why this
scale is not relevant for standard-model fermions.  Our principal results,
which we elaborate upon in the body of the paper, may be summarized as follows:
\begin{itemize}

\item In the standard model,\footnote{Throughout this paper, the standard
model refers to the $SU(2)_L\times U(1)_Y$ gauge theory spontaneously broken
by the vacuum-expectation value of a Higgs doublet field, including all terms
of dimension four and less.  We regard terms of dimension greater than four as
beyond the standard model.} there is {\em no} scale of fermion mass
generation. The Higgs-boson mass is the scale of electroweak symmetry
breaking, but it is not the scale of fermion mass generation.

\item The upper bound on the scale of fermion mass generation depends on the
dimensionality of the interaction responsible for generating the fermion mass.
The upper bound is proportional to $(v^{d-3}/m_f)^{1/(d-4)}$, where $d > 4$ is
the dimensionality of the interaction.  This is less than $\Lambda_f$ except
for $d=5$, when it is equal to it.  For $d\le 4$, there is no upper bound on
the scale of fermion mass generation.

\item If electroweak symmetry breaking is not driven
via the vacuum-expectation value of a Higgs field, one cannot derive an upper
bound on the scale of fermion mass generation by considering
fermion-antifermion scattering into longitudinal weak vector bosons.

\end{itemize}

The remainder of the paper is organized as follows.  In Section~\ref{EWSB} we
revisit the upper bound on the scale of electroweak symmetry breaking,
$\Lambda_{EWSB}$, in order to prepare for the discussion of the scale of
fermion mass generation. In Section~\ref{standard} we show that there is no
upper bound on the scale of fermion mass generation, by considering
fermion-antifermion scattering into a large number of longitudinal weak vector
bosons.  In Section~\ref{standardmodel} we show that there is no scale of
fermion mass generation in the standard model with a Higgs boson.  In
Section~\ref{neutrino} we show that the upper bound on the scale of
Majorana-neutrino mass generation is proportional to $v^2/m_\nu$ and can be
naturally saturated in explicit models. In Section~\ref{nonstandard} we
consider an extension of the standard model with fermions in non-standard
representations of the gauge group, such that their masses arise via
higher-dimension interactions. This allows us to study the upper bound on the
scale of fermion mass generation in a more general setting. In
Section~\ref{higherdimension} we return to the standard model with the usual
fermion content, but including higher-dimension interactions.  The
two-Higgs-doublet model, in the limit that one doublet is much heavier than
the weak scale, provides a specific example of such a theory and allows us to
recover the results of Ref.~\cite{Jager:1998va} in a simple way. In
Section~\ref{strong} we consider models without a Higgs field.  We summarize
our conclusions in Section~\ref{conclusions}.

\section{Scale of electroweak symmetry breaking}\label{EWSB}

We begin with the well-established upper bound on the scale of electroweak
symmetry breaking.  Consider an $SU(2)_L\times U(1)_Y$ Yang-Mills gauge theory.
The weak vector bosons are massless due to the gauge symmetry.  Now add a bare
mass for the $W$ and $Z$ bosons,
\begin{equation}
{\cal L} = M_W^2W^{+\mu} W^-_\mu + \frac{1}{2}\frac{M_W^2}{\cos^2\theta_W}Z^\mu
Z_\mu\;, \label{Wmass}
\end{equation}
where the relation $M_W^2 = M_Z^2\cos^2\theta_W$ is made explicit.  These terms
violate the gauge symmetry, so one should question why it is legitimate to add
them. The answer is that these terms correspond to the unitary-gauge
expression of an effective Lagrangian in which the gauge symmetry is
nonlinearly realized,
\begin{equation}
{\cal L} = \frac{v^2}{4}{\rm Tr}(D^\mu \Sigma)^\dagger D_\mu \Sigma\;,
\label{Wmassprime}
\end{equation}
where $D^\mu\Sigma = \partial^\mu\Sigma + i(g/2)\sigma\cdot W^\mu \Sigma-i
(g^\prime/2)\Sigma\sigma^3B^\mu$ and $\Sigma = \exp(i\sigma\cdot\pi/v)$
contains the Goldstone bosons $\pi^i$ of the spontaneously-broken gauge
symmetry~\cite{Callan:1969sn,Appelquist:1980vg}. This effective
field theory is valid below the scale of electroweak symmetry breaking, but not
above. One may then calculate the scale at which this effective field theory
breaks down, $\Lambda_{EWSB}$. The theory that subsumes this effective field
theory and contains the physics of electroweak symmetry breaking must occur at
or below this scale.  Thus $\Lambda_{EWSB}$ represents an upper bound on the
scale of electroweak symmetry breaking.

The scale at which the effective field theory breaks down may be calculated
using unitarity.  The zeroth-partial-wave ($J=0$) elastic scattering amplitude
for longitudinal weak vector bosons is proportional to $s/v^2$, where $s$ is
the square of the center-of-mass energy and $v= (\sqrt 2 G_F)^{-1/2}$ is the
weak scale.\footnote{In the standard model, $v$ is the vacuum-expectation
value of the Higgs field. However, there is no Higgs field in the effective
field theory of massive weak vector bosons.  In the effective theory, $v$ is
defined by Eq.~(\ref{Wmassprime}).} Applying the elastic unitarity condition
$|{\rm Re}\;a^0_0| \le 1/2$ to the $J=0$, $I=0$ partial-wave
amplitude\footnote{Weak isospin, $I$, is an approximate global $SU(2)$
symmetry of the effective field theory and is exact in the limit $\cos\theta_W
= 1$. This symmetry is manifest in this limit by the weak-vector-boson masses,
Eq.~(\ref{Wmass}), where $W^+,Z,W^-$ form an isotriplet.  It is also manifest
in Eq.~(\ref{Wmassprime}) in this limit ($g^\prime = 0$), where the $\pi^i$
form an isotriplet.} yields the energy at which the effective field theory
breaks down \cite{Chanowitz:1985hj,Marciano:1989ns},
\begin{equation}
\Lambda_{EWSB} \equiv \sqrt{8\pi}v \approx 1\;{\rm TeV}\;.
\end{equation}
This is the upper bound on the scale of electroweak symmetry breaking.

In the standard model at energies above the Higgs-boson mass, the elastic
scattering amplitude for longitudinal weak vector bosons receives an additional
contribution from the exchange of the Higgs boson.  This contribution cancels
the term proportional to $s/v^2$, leaving behind terms that approach a constant
at high energy.  Thus the effective field theory of massive weak vector bosons
is subsumed by a deeper theory containing a Higgs boson.

At energies above the Higgs mass, the Lagrangian describing the theory has a
linearly-realized $SU(2)_L\times U(1)_Y$ gauge invariance, unlike the effective
field theory of massive weak vector bosons that operates below the Higgs mass.
The Lagrangian of Eq.~(\ref{Wmassprime}) is replaced by
\begin{equation}
{\cal L} = (D^\mu \phi)^\dagger D_\mu \phi - \lambda (\phi^\dagger
\phi-v^2/2)^2\;, \label{Higgs}
\end{equation}
where $\phi$ is the Higgs doublet field.  One may recover the effective field
theory of massive weak vector bosons at energies less than the Higgs mass,
Eq.~(\ref{Wmassprime}), by integrating out the Higgs-boson field, $h$,
contained in the Higgs doublet field, $\phi = \Sigma \,(0, (h+v)/\sqrt
2)$.

The above considerations lead us to the following definition: {\em The scale
of electroweak symmetry breaking is the minimum energy at which the Lagrangian
has a linearly-realized $SU(2)_L\times U(1)_Y$ gauge invariance.}  In the
standard model, the Higgs-boson mass is the scale of electroweak symmetry
breaking.

The Higgs-boson mass is proportional to $\sqrt\lambda v$, where $\lambda$ is
the Higgs-field self coupling in Eq.~(\ref{Higgs}).  Since the coupling is
bounded to be at most of order $4\pi$, the upper bound on the Higgs mass is
approximately $\sqrt{4\pi}v$ \cite{Dashen:1983ts}. This is derived by
requiring that the Higgs mass be less than the ultraviolet cutoff of the
theory. The upper bound on the Higgs-boson mass is parametrically the same as
the upper bound on the scale of electroweak symmetry breaking, $\Lambda_{EWSB}
\equiv \sqrt{8\pi}v$, so the Higgs mass can saturate this bound within a
factor of order unity. A detailed analysis shows that the upper bound on the
Higgs mass is approximately 600 GeV \cite{Luscher:1988gc}.

If there is no Higgs boson, then the effective field theory of massive weak
vector bosons simply ceases to provide a valid description of nature above
$\Lambda_{EWSB}$.  In particular, the theory that describes physics above
$\Lambda_{EWSB}$ will not contain longitudinal weak vector bosons as
weakly-coupled degrees of freedom.  The standard model (and extensions thereof
that decouple \cite{Appelquist:1975tg} when the mass of the additional physics
is taken to infinity) is the unique theory that contains longitudinal weak
vector bosons as weakly-coupled degrees of freedom above $\Lambda_{EWSB}$
\cite{Cornwall:1973tb,Cornwall:1974km,LlewellynSmith:1973ey}.  Since a theory
of Goldstone bosons $\Sigma$, but no Higgs boson, does not possess linearly-realized
gauge symmetry, the scale of electroweak symmetry breaking typically
saturates $\Lambda_{EWSB}$ in such models.
We consider strongly-coupled models in Section~\ref{strong}.

\section{Scale of fermion mass generation}
\label{standard}

\begin{figure}[t]
\begin{center}
\vspace*{0cm} \hspace*{0cm} \epsfxsize=14cm \epsfbox{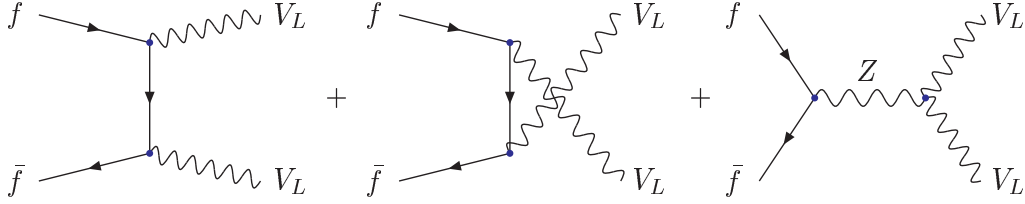} \vspace*{0cm}
\caption{Feynman diagrams that contribute to the amplitude for $f\bar f \to
V_LV_L$ in unitary gauge.  The middle diagram is absent if $V=W$; the last
diagram is absent if $V=Z$.} \label{ffVV}
\end{center}
\end{figure}

The upper bound on the scale of fermion mass generation derived by Appelquist
and Chano\-witz is based on a calculation of $f_\pm \bar f_\pm \to V_LV_L$
(where $V_L$ is a longitudinal weak vector boson and the subscripts on the
fermion and antifermion indicate their helicities), as shown in Fig.~\ref{ffVV}
\cite{Appelquist:1987cf}. The fermion mass is introduced via a bare mass term
in the Lagrangian,
\begin{equation}
{\cal L} = -m_f{\overline f_L}f_R+H.c.\;, \label{fmass}
\end{equation}
where the subscripts indicate chirality.  This term violates the gauge symmetry
since, in the standard model, $f_L$ and $f_R$ transform differently under
$SU(2)_L\times U(1)_Y$ gauge transformations.  Actually, Eq.~(\ref{fmass}) is
the unitary-gauge expression of a Lagrangian in which the gauge symmetry is
nonlinearly realized,
\begin{equation}
{\cal L} = -m_f{\overline F_L} \Sigma {0 \choose 1} f_R+H.c.\;,
\label{fmassprime}
\end{equation}
where $F_L$ is an $SU(2)_L$-doublet fermion field whose lower component is
$f_L$.  Since the fermion mass is not introduced via a Yukawa coupling to the
Higgs field, there is no diagram corresponding to the exchange of a Higgs
boson in the $s$-channel, as there would be in the standard model. The
resulting amplitude is proportional to the fermion mass and grows linearly
with energy. Applying the inelastic unitarity condition $|a_0^0| \le 1/2$ to
the $J=0$, $I=0$, spin-zero, color-singlet amplitude for $f_\pm \bar f_\pm \to
V_LV_L$ leads to an upper bound on the scale of fermion mass generation
\cite{Appelquist:1987cf,Marciano:1989ns}
\begin{equation}
\Lambda_f \equiv \frac{8\pi v^2}{\sqrt{3N_c}m_f}\;, \label{lambdaf}
\end{equation}
where $N_c=3$ for quarks and unity for leptons.

However, Eq.~(\ref{lambdaf}) is not the strongest upper bound that one can
derive, given the above framework.  By considering $f_\pm\bar f_\pm \to
V_L\cdots V_L$, with $n$ particles in the final state, one obtains an upper
bound on the scale of fermion mass generation proportional to
$(v^n/m_f)^{1/(n-1)}$. For arbitrarily large $n$, one obtains an upper bound
arbitrarily close to the weak scale $v$ for any value of $m_f$.  We first
derive this result, then discuss its implications.

The easiest way to derive this result is to consider the theory in the limit
that the weak gauge coupling goes to zero, with $v$ fixed.  In this limit the
weak vector bosons become massless, and the longitudinal weak vector bosons
are represented by the Goldstone bosons $s^\pm, \chi$ contained in the field
$\Sigma = \exp(i\sigma\cdot\pi/v)$, where $s^\pm=-(\pi^1 \mp i\pi^2)/\sqrt 2,
\chi = -\pi^3$. The terms that grow with energy in the amplitudes are
independent of the weak gauge coupling, so they survive in this limit. Thus
the high-energy behavior of amplitudes with longitudinal weak vector bosons in
the final state may be obtained from the amplitudes with the vector bosons
replaced with the corresponding Goldstone bosons [times a factor of $i$ ($-i$)
for each outgoing (incoming) longitudinal weak vector boson]. This is the
Goldstone-boson equivalence theorem
\cite{Chanowitz:1985hj,Cornwall:1974km,Vayonakis:1976vz,Lee:1977eg}.\footnote{The
Goldstone-boson equivalence theorem is actually more general, being valid for
finite weak gauge coupling \cite{Bagger:1990fc}.}

\begin{figure}[t]
\begin{center}
\vspace*{0cm} \hspace*{0cm} \epsfxsize=4cm \epsfbox{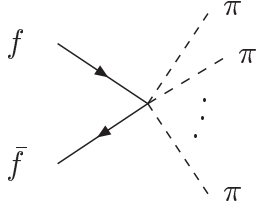} \vspace*{0cm}
\caption{Feynman diagram for the interaction of a fermion with $n$ Goldstone
bosons.} \label{ffpipi}
\end{center}
\end{figure}

The fermion interacts with the Goldstone bosons via the interaction of
Eq.~(\ref{fmassprime}).  Expanding the $\Sigma$ field in powers of the
Goldstone-boson fields, we obtain an interaction such as that shown in
Fig.~\ref{ffpipi}, with $n$ external Goldstone bosons. The Feynman rule for
this interaction is proportional to $m_f/v^n$. The amplitude for $f_\pm\bar
f_\pm\to \pi\cdots \pi$ is therefore proportional to $m_f\sqrt s/v^n$.  The
relevant unitarity condition on this inelastic amplitude is
\begin{equation}
\sigma_{inel}(2\to n) \le \frac{4\pi}{s}\;, \label{unitarity}
\end{equation}
where $\sigma_{inel}(2\to n)$ is the total cross section for $f_\pm\bar
f_\pm\to \pi\cdots \pi$. This condition is derived in Appendix A.  Since the
phase space for an $n$-particle final state is proportional to $s^{n-2}$ at
high energies, one finds that the unitarity condition, Eq.~(\ref{unitarity}),
is violated at an energy proportional to $(v^n/m_f)^{1/(n-1)}$, as stated
above.

We see that $f_\pm\bar f_\pm \to V_L\cdots V_L$, with $n>2$ particles in the
final state, leads to a stronger upper bound than Eq.~(\ref{lambdaf}), which
is based on the case $n=2$.  Thus the Appelquist-Chanowitz bound is subsumed
by this stronger bound, which is of order the weak scale, $v$, for $n$ large,
independently of $m_f$.  Since we already know that there must be new physics
at the weak scale, namely the physics of electroweak symmetry breaking, the
consideration of fermion-antifermion scattering into longitudinal weak vector
bosons does not reveal an additional scale.  This claim is supported by the
fact the upper bound is independent of the fermion mass.  Thus there is no
upper bound on the scale of fermion mass generation.

\section{Standard model}
\label{standardmodel}

The derivation in the previous section of $f_\pm\bar f_\pm \to V_L\cdots V_L$,
with $n$ particles in the final state, tacitly assumes that the longitudinal
weak vector bosons are weakly-coupled degrees of freedom. As discussed in
Section~\ref{EWSB}, this is not true in general above $\Lambda_{EWSB}\approx
\sqrt{8\pi}v$. In order to justify the calculation of $f_\pm\bar f_\pm \to
V_L\cdots V_L$ above $\Lambda_{EWSB}$, one must specify the mechanism of
electroweak symmetry breaking such that the longitudinal weak vector bosons
remain weakly-coupled degrees of freedom above $\Lambda_{EWSB}$.  The unique
theory that contains longitudinal weak vector bosons as weakly-coupled degrees
of freedom to arbitrarily-high energies is the standard model, with a Higgs
boson \cite{Cornwall:1973tb,Cornwall:1974km,LlewellynSmith:1973ey}.  In this
section we consider the scale of fermion mass generation in the standard model.

First consider the model envisioned in Ref.~\cite{Appelquist:1987cf}, in which
the weak-vector-boson masses are generated via an explicit model of spontaneous
symmetry breaking, but fermions are given bare masses.  As an example of this,
one could imagine the standard Higgs model, but with the fermion Yukawa
interactions replaced by bare fermion masses, Eq.~(\ref{fmass}).  However,
even in this scenario, the considerations of the previous section continue to
apply. The calculation of $f_\pm\bar f_\pm \to V_L\cdots V_L$, with $n$
particles in the final state, continues to violate unitarity at the scale of
electroweak symmetry breaking for large $n$. Thus unitarity of this process
does not reveal an additional scale beyond that of electroweak symmetry
breaking.

\begin{figure}[t]
\begin{center}
\vspace*{0cm} \hspace*{0cm} \epsfxsize=4.5cm \epsfbox{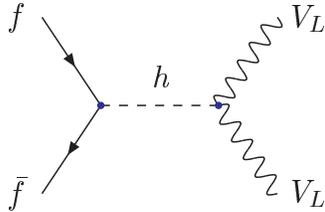} \vspace*{0cm}
\caption{Additional diagram involving the exchange of a Higgs boson that
contributes to the amplitude for $f\bar f \to V_LV_L$. This diagram cancels
the terms that grow with energy resulting from the diagrams in
Fig.~\ref{ffVV}.} \label{ffHiggs}
\end{center}
\end{figure}

The theory that is valid above the scale of electroweak symmetry breaking
necessarily has a linearly-realized gauge invariance.  Thus the fermion mass,
Eq.~(\ref{fmassprime}), must be described by a Yukawa interaction\footnote{This
interaction may be supplemented by additional interactions of dimension
greater than four that also contribute to the fermion mass.  We consider this
possibility in Section~\ref{higherdimension}.}
\begin{equation}
{\cal L} = -y_f{\overline F_L} \phi f_R+H.c.\;. \label{Yukawa}
\end{equation}
This Lagrangian contains a Yukawa interaction of the fermion with the Higgs
boson and yields the diagram in Fig.~\ref{ffHiggs}.  This diagram, when added
to the diagrams in Fig.~\ref{ffVV}, cancels the term that grows linearly with
energy, leaving behind terms that fall like an inverse power of energy at high
energy.  A similar cancellation occurs for all processes of the type
$f_\pm\bar f_\pm \to V_L\cdots V_L$.

It is tempting to identify the scale of fermion mass generation with the
energy at which the amplitude for $f_\pm \bar f_\pm \to V_LV_L$ ceases to grow
with energy, namely the Higgs mass.  However, the Higgs mass is the scale of
electroweak symmetry breaking, not the scale of fermion mass generation.  The
reason the amplitude for $f_\pm \bar f_\pm \to V_LV_L$ grows with energy below
the Higgs mass is because the fermion mass is described in a theory with a
nonlinearly-realized gauge invariance, Eq.~(\ref{fmassprime}).  Above the Higgs
mass, the amplitude for $f_\pm \bar f_\pm \to V_LV_L$ falls off with energy and
unitarity is respected at all energies. Thus, in the standard model there is
{\em no} scale associated with fermion mass generation.  We will support this
claim by considering extensions of the standard model in which there {\em is} a
well-defined scale of fermion mass generation. These models are discussed in
Sections~\ref{neutrino} and \ref{nonstandard}.

A possible way to circumvent the above arguments is to introduce a Higgs
doublet field, such that longitudinal weak vector bosons are weakly coupled
above the weak scale, but to forbid the Higgs field from coupling to fermions.
This can be arranged, for example, by imposing the discrete symmetry $\phi\to
-\phi$.  However, this also has the consequence of forbidding a
gauge-invariant mass for the fermion, so the scale of fermion mass generation
is moot.  One might also consider a model with two Higgs doublets where only
one doublet couples to fermions.  Such a model is discussed in
Section~\ref{higherdimension}.

In this section we have argued that there is no scale of fermion mass
generation in the standard model. However, Yukawa couplings are not
asymptotically free in general, so the energy at which a Yukawa coupling
becomes strong also indicates an upper bound on the scale of fermion mass
generation.  In the standard model, only the top-quark Yukawa coupling is not
asymptotically free; all other Yukawa couplings are asymptotically free by
virtue of the fermion's gauge interactions.  The top-quark's Yukawa coupling is
sufficiently large that it eventually overwhelms the gauge interactions,
causing it to become strong at high energies.  However, for $m_t=175$ GeV, the
energy at which the top-quark's Yukawa coupling becomes strong is many orders
of magnitude above the Planck scale and is therefore irrelevant. If a quark of
mass in excess of about 225 GeV existed, its Yukawa coupling would become
strong below the grand-unification scale
\cite{Maiani:1978cg,Cabibbo:1979ay,Pendleton:1981as,Bardeen:1990ds}.

\section{Majorana neutrinos}\label{neutrino}

Neutrinos are exactly massless in the standard model.  However, recent
observations of neutrino oscillations indicate that neutrinos have a small
mass.  We assume that  neutrino masses are Majorana, unlike the other known
fermions, which carry electric charge and are therefore forbidden to have
Majorana masses.  If there is no $SU(2)_L\times U(1)_Y$-singlet fermion field
$\nu_R$ in nature, then neutrino masses are necessarily Majorana. However,
even if such a field exists, the gauge symmetry allows the Majorana mass term
${\cal L} = -(M_R/2) \nu_R^T C \nu_R+H.c.$ for this field, and there is no reason
why this mass should be small. Other known fermions acquire a mass only
after  $SU(2)_L\times U(1)_Y$  is broken, and thus their masses are
of order the weak scale, $v$, or less. Since a Majorana mass for the $\nu_R$
field is not protected by the gauge symmetry, it is natural to assume that it
would be much greater than the weak scale \cite{Georgi:1979md}. So even if the
$\nu_R$ field exists, it is likely to be heavy, in which case the light
neutrinos are Majorana fermions.

\begin{figure}[t]
\begin{center}
\vspace*{0cm} \hspace*{0cm} \epsfxsize=14cm \epsfbox{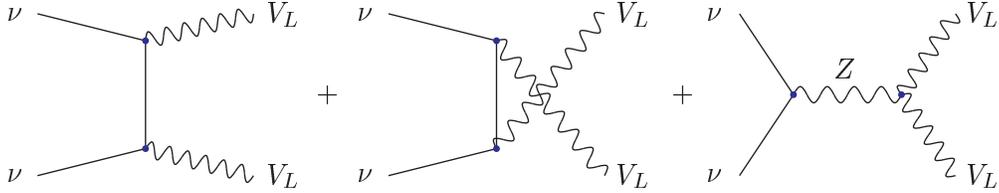} \vspace*{0cm}
\caption{Feynman diagrams that contribute to the amplitude for $\nu\nu\to
V_LV_L$ in unitary gauge. The last diagram is absent if $V=Z$.} \label{nunuVV}
\end{center}
\end{figure}

We have recently shown that an upper bound on the scale of Majorana-neutrino
mass generation may be derived by considering the process $\nu_\pm\nu_\pm\to
V_LV_L$, as shown in Fig.~\ref{nunuVV} \cite{Maltoni:2001iq}. This bound is
similar to the Appelquist-Chanowitz bound on Dirac-fermion mass generation,
Eq.~(\ref{lambdaf}), which is invalid for standard-model fermions, as we have
argued in the previous two sections.  Here we reconsider the upper bound on the
scale of Majorana-neutrino mass generation and show that it is valid.

As with the case of Dirac fermions, the upper bound on the scale of
Majorana-neutrino mass generation was obtained by considering the process
$\nu_\pm\nu_\pm\to V_LV_L$ in the absence of any diagrams involving the
exchange of a Higgs boson.  This is because the Majorana-neutrino mass was
introduced via a bare mass term,
\begin{equation}
{\cal L} = -\frac{1}{2}m_\nu\nu_L^TC\nu_L+H.c.\;,
\label{numass}
\end{equation}
where $C$ is the charge-conjugation matrix.  However, by considering
$\nu_\pm\nu_\pm\to V_L \cdots V_L$, with $n$ particles in the final state, one
finds that unitarity is violated at the weak scale, $v$, for $n$ large,
independently of the neutrino mass.  This is analogous to the situation for
Dirac fermions discussed in Section~\ref{standard}.  Thus there is no
additional upper bound, beyond that on electroweak symmetry breaking, implied
by considering Majorana neutrinos scattering into longitudinal weak vector
bosons when the neutrino mass is introduced via a bare mass term,
Eq.~(\ref{numass}).

In order to discover a new scale from the consideration of $\nu_\pm\nu_\pm\to
V_LV_L$, one must allow the neutrino to acquire a mass by coupling to the
Higgs boson.  This has two consequences.  First, the longitudinal weak vector
bosons remain weakly coupled up to arbitrarily high energies, justifying the
calculation of the diagrams in Fig.~(\ref{nunuVV}).  Second, the process
$\nu_\pm\nu_\pm\to V_L \cdots V_L$, with $n$ particles in the final state,
does not lead to a stronger bound than the case with $n=2$.  If the neutrino
instead acquires its mass some other way, then the considerations of this
section do not apply.  This case is treated in Section~\ref{strong}.

\begin{figure}[t]
\begin{center}
\vspace*{0cm} \hspace*{0cm} \epsfxsize=4.5cm \epsfbox{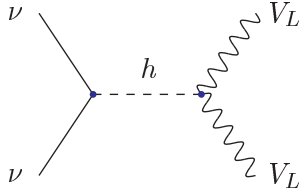}
\vspace*{0cm} \caption{Additional diagram involving the exchange of a Higgs
boson that contributes to the amplitude for $\nu\nu\to V_LV_L$. This diagram
cancels the term that grows with energy resulting from the diagrams in
Fig.~\ref{nunuVV} if $V=W$, but not if $V=Z$.} \label{nunuHiggs}
\end{center}
\end{figure}

Above the scale of electroweak symmetry breaking, the Majorana-neutrino mass
must be described by a gauge-invariant term in the Lagrangian.   In the Higgs
model, the lowest-dimension term available is the dimension-five interaction
\cite{Weinberg:1979sa}
\begin{equation}
{\cal L} = \frac{c}{M}(L^T\epsilon\phi)C(\phi^T\epsilon L)+H.c.\;,
\label{nuYukawa}
\end{equation}
where $L=(\nu_L,\ell_L)$ is an $SU(2)_L$ doublet containing the left-chiral
neutrino and charged-lepton fields and $\epsilon\equiv i\sigma_2$.  We will
show that the scale $M$ may be interpreted as the scale of Majorana-neutrino
mass generation; $c$ is a dimensionless constant. This term gives rise to a
Majorana-neutrino mass $m_\nu=cv^2/M$ when the neutral component of the Higgs
field acquires a vacuum-expectation value $\langle\phi^0\rangle=v/\sqrt 2$. It
also yields a Yukawa coupling of the Majorana neutrino to the Higgs boson,
thereby generating the additional contribution to the amplitude
$\nu_\pm\nu_\pm\to V_LV_L$ shown in Fig.~\ref{nunuHiggs}. However, this
diagram does not cancel the terms that grow with energy,\footnote{To be
precise, the Higgs-exchange diagram {\em does} cancel the term that grows with
energy in $\nu_\pm\nu_\pm\to W_L^+W_L^-$; however, it does not cancel this
term in $\nu_\pm\nu_\pm\to Z_LZ_L$, nor in $\ell_-\nu_-\to W_L^-Z_L$ or
$\ell_-\ell_-\to W_L^-W_L^-$ \cite{Maltoni:2001iq}.} in contrast to the case of
standard-model Dirac fermions.  Thus the upper bound on the scale of
Majorana-neutrino mass generation derived in Ref.~\cite{Maltoni:2001iq} is
parametrically correct, although it did not include the contribution from the
Higgs-exchange diagram in Fig.~\ref{nunuHiggs}.

\begin{figure}[t]
\begin{center}
\vspace*{0cm} \hspace*{0cm} \epsfxsize=16cm \epsfbox{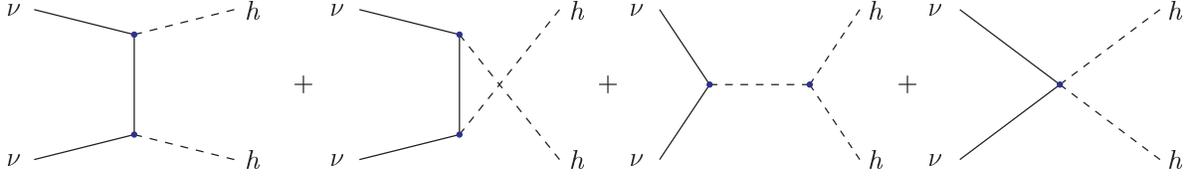} \vspace*{0cm}
\caption{Feynman diagrams that contribute to the amplitude for $\nu\nu\to hh$.
Only the last diagram grows with energy.}  \label{nunuHH}
\end{center}
\end{figure}

Since the Higgs boson is present at energies above the scale of electroweak
symmetry breaking, one finds that there is another amplitude that grows with
energy, $\nu_\pm\nu_\pm\to hh$, as shown in Fig.~\ref{nunuHH}.\footnote{The
amplitudes for $\nu_\pm\nu_\pm\to Z_Lh$ and $\ell_-\nu_-\to W_L^-h$ also grow
with energy.} Only the last diagram contributes to the term that grows with
energy, yielding the zeroth-partial-wave amplitude (for $\sqrt s\gg m_\nu,m_h$)
\begin{equation}
a_0\left(\frac{1}{\sqrt 2}\nu_\pm\nu_\pm\to\frac{1}{\sqrt 2}hh\right) \sim
\mp\frac{c\sqrt s}{16\pi M} \sim \mp\frac{m_\nu\sqrt s}{16\pi v^2}\;,
\end{equation}
where the relation $m_\nu = cv^2/M$ was used to obtain the final expression.
This process grows with energy because the interaction responsible for the
last diagram in Fig.~\ref{nunuHH}, Eq.~(\ref{nuYukawa}), has a coefficient with
dimensions of an inverse power of mass. In contrast, the processes involving
longitudinal weak vector bosons in the final state grow with energy due to the
longitudinal polarization vectors, $\epsilon^\mu \approx p^\mu/M_V$ (for
$p^0\gg M_V$).

However, there is a sense in which {\em all} processes that grow with energy
are related to the dimension-five interaction of Eq.~(\ref{nuYukawa}). This
can be made manifest by using the Goldstone-boson equivalence theorem, where
the Goldstone bosons $s^\pm,\chi$ are contained in the Higgs doublet,
$\phi=(-is^+,(h+v+i\chi)/\sqrt 2)$. The terms that grow with energy in the
Goldstone-boson amplitudes all come from the interaction of
Eq.~(\ref{nuYukawa}), such as the last diagram in Fig.~\ref{nunuHH}.  It is in
this sense that all processes that grow with energy are related to this
dimension-five interaction.

The high-energy behavior of the amplitudes that grow with energy are collected
in Appendix B.  The strongest upper bound on the scale of Majorana-neutrino
mass generation is obtained by applying the inelastic unitarity condition
$|a_0| \le 1/2$ to the amplitude\footnote{The same bound may also be obtained
by considering the amplitude
\begin{eqnarray} a_0\left(\frac{1}{2}(\nu_{i+}\nu_{i+} + \nu_{i-}\nu_{i-}) \to
Z_Lh\right) \sim -\frac{m_{\nu_i} \sqrt s}{8\pi
v^2}\;.\nonumber
\end{eqnarray}}
\begin{equation}
a_0\left(\frac{1}{2}(\nu_{i+}\nu_{i+} - \nu_{i-}\nu_{i-}) \to
\frac{1}{2}(Z_LZ_L+hh)\right) \sim -\frac{m_{\nu_i} \sqrt s}{8\pi
v^2}\;.\label{a0neutrino}
\end{equation}
This yields the upper bound on the scale of Majorana-neutrino mass generation
\begin{equation}
\Lambda_{Maj} \equiv \frac{4\pi v^2}{m_{\nu}}\;. \label{nubound}
\end{equation}
This equation supersedes Eq.~(10) of Ref.~\cite{Maltoni:2001iq}.
The upper bounds on $\Lambda_{Maj}$ implied by a variety of neutrino-oscillation
experiments are listed in Table~\ref{massdiffs}.

\begin{table}[t]
\caption{Neutrino mass-squared differences from a variety of neutrino
oscillation experiments, and their interpretations.  They imply a lower
bound of $m_\nu\ge\sqrt{\Delta m^2}$ on the mass of one of the two
participating neutrino species. The last column lists the corresponding
upper bounds on $\Lambda_{Maj}$, Eq.~(\ref{nubound}),
which is the upper bound on the scale of Majorana-neutrino mass generation.}
\begin{center}
\begin{tabular}{lll|c} \hline \hline &&&\\[-.4cm]
Experiment&Fav'd Channel(s)&$\Delta m^2$ ($\rm eV^2$)
&$\Lambda_{Maj}$ $({\rm GeV})<$\\ \hline &&&\\[-.4cm]
LSND \cite{Aguilar:2001ty}&$\bar\nu_\mu\to\bar\nu_e$&$0.2-2.0$&$1.7\times10^{15}$\\
Atmospheric \cite{Toshito:2001dk}&$\nu_\mu\to\nu_\tau$&$(1.6-4)
\times10^{-3}$&$1.9\times10^{16}$\\
Solar \cite{Bahcall:2001hv} & & & \\
\quad MSW (LMA) &$\nu_e\to\nu_\mu$ or $\nu_\tau$&
$(2-10)\times10^{-5}$&$1.7\times10^{17}$\\
\quad MSW (SMA)&$\nu_e\to$ anything &$(3-8)\times10^{-6}$
&$4.4\times 10^{17}$\\
\quad MSW (LOW)&$\nu_e\to\nu_\mu$ or $\nu_\tau$ &$7.6\times10^{-8}$
&$2.8\times 10^{18}$\\
\quad Vacuum&$\nu_e\to$ anything&$1.4\times10^{-10}$&$6.4\times10^{19}$\\
\quad Just So$^2$&$\nu_e\to$ anything&$5.5\times10^{-12}$&$3.2\times10^{20}$\\
\hline \hline
\end{tabular}
\end{center}
\label{massdiffs}
\end{table}

In Ref.~\cite{Maltoni:2001iq} we discussed two models that can saturate the
upper bound on the scale of Majorana-neutrino mass generation, Eq.~(\ref{nubound}):
the ``see-saw'' model and a Higgs-triplet model.  We first review the see-saw model
\cite{GRS,Mohapatra:1980ia}.  In this model, the dimension-five interaction of
Eq.~(\ref{nuYukawa}) is replaced by the renormalizable interactions
\begin{equation}
{\cal L} = -y_D \overline L\epsilon \phi^*\nu_R-\frac{1}{2}M_R \nu_R^T C
\nu_R+H.c.\;, \label{seesaw}
\end{equation}
where $\nu_R$ is an $SU(2)_L\times U(1)_Y$-singlet fermion field.  This field
has a Majorana mass term allowed by the gauge symmetry, so it is
natural to expect that $M_R \gg v$. The first term yields a Dirac mass of $m_D
= y_Dv/\sqrt 2$. The mass eigenstates of this model are a light Majorana
neutrino $\nu\approx \nu_L$, of mass $m_\nu\approx m_D^2/M_R$, and a heavy
Majorana neutrino $N\approx \nu_R$, of approximate mass $M_R$. The fact that
$m_\nu \ll m_D$ provides an attractive explanation for why neutrinos are so
much lighter than the other known (Dirac) fermions. At energies above the mass
of the heavy neutrino, $M_R$, the Feynman diagrams for $\nu_\pm\nu_\pm \to
Z_LZ_L$ in Figs.~\ref{nunuVV} and \ref{nunuHiggs} are augmented by diagrams in
which the heavy neutrino is exchanged in the $t$- and $u$-channels.  These
diagrams cancel all terms that grow with energy.\footnote{Similarly, all terms
that grow with energy are cancelled in $\ell_-\nu_-\to W_L^-Z_L$, etc.} The
process $\nu_\pm\nu_\pm\to hh$ also ceases to grow with energy, because the
last diagram in Fig.~\ref{nunuHH}, which was responsible for the term that
grows with energy, is not present.  It is replaced by diagrams, similar to the
first two diagrams in that figure, with the exchange of $N$ in the $t$- and
$u$-channels. Thus the scale of Majorana-neutrino mass generation in the
see-saw model is the mass of the heavy neutrino, $M_R$. This is because the
Lagrangian above $M_R$, Eq.~(\ref{seesaw}), is renormalizable.

Below $M_R$, one integrates out the field $\nu_R$ and obtains the
nonrenormalizable interaction of Eq.~(\ref{nuYukawa}), with
$c/M=-y_D^2/2M_R$.  Thus we associate the scale $M$ with $M_R$, which is the
scale of Majorana-neutrino mass generation in this model, and $c=-y_D^2/2$.
The mass of the heavy neutrino, $M_R\approx m_D^2/m_\nu \approx
y_D^2v^2/2m_\nu$, saturates (within a factor of order unity) the upper bound on
the scale of Majorana-neutrino mass generation, Eq.~(\ref{nubound}), when the
Yukawa coupling takes its largest allowed value, $y_D \ltap \sqrt{8\pi}$
\cite{Chanowitz:1979mv,Einhorn:1986za,Lee:1990xq}.

\begin{figure}[t]
\begin{center}
\vspace*{0cm} \hspace*{0cm}
\epsfxsize=14cm \epsfbox{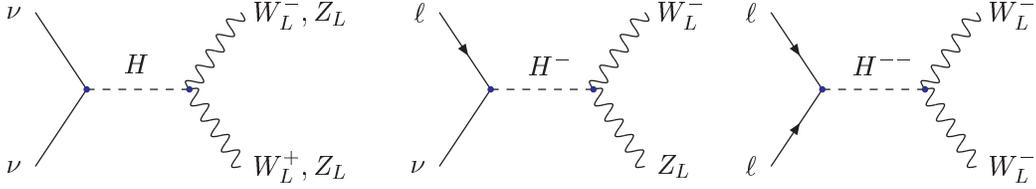} \vspace*{0cm} \caption{Additional
diagrams that contribute when the Majorana neutrino acquires its mass via a
coupling to an SU(2)$_L$-triplet Higgs field.} \label{ffVVtriplet}
\end{center}
\end{figure}

The Higgs-triplet model
\cite{Cheng:1980qt,Gelmini:1981re,Mohapatra:1981yp,Georgi:1981pg,Ma:1998dx}
introduces an SU(2)$_L$-triplet, $Y=1$ Higgs field, $\Phi^i$, and the
renormalizable interaction
\begin{equation}
{\cal L} = -y_M L^T\epsilon\sigma^iCL\Phi^i+H.c.\;, \label{htriplet}
\end{equation}
which replaces the dimension-five interaction of Eq.~(\ref{nuYukawa}).  The
usual Higgs doublet field is also present in the model.  The vacuum-expectation
value of the Higgs triplet field must be much less than the weak scale, because
the relation $M_W^2 \simeq M_Z^2\cos^2\theta_W$, which is satisfied
experimentally, is obtained if the weak bosons acquire their mass dominantly
from the vacuum-expectation value of an SU(2)$_L$ doublet, but not a triplet.
The interaction of Eq.~(\ref{htriplet}) generates a small Majorana-neutrino
mass, $m_\nu = 2y_Mu$, when the neutral component of the Higgs field,
$\Phi^0=(\Phi^1+i\Phi^2)/\sqrt 2$, acquires a small vacuum-expectation value
$\langle \Phi^0\rangle = u/\sqrt 2$. This model contains three neutral
scalars, one singly-charged scalar, and one doubly-charged scalar. The term of
Eq.~(\ref{htriplet}) gives rise to new interactions that yield the additional
Feynman diagrams in Fig.~\ref{ffVVtriplet} involving these Higgs scalars in
the intermediate state.\footnote{We impose CP conservation in this model, in
which case one of the neutral scalars is CP odd and does not contribute to the
amplitudes.} The first diagram cancels the terms that grow with energy in
$\nu_\pm\nu_\pm\to V_LV_L$, the second diagram cancels the term that grows
with energy in $\ell_-\nu_-\to W_L^-Z_L$, and the third diagram cancels the
term that grows with energy in $\ell_-\ell_-\to W_L^-W_L^-$.\footnote{Terms
that grow with energy are similarly cancelled in $\nu_\pm\nu_\pm\to Z_Lh$ and
$\ell_-\nu_-\to W_L^-h$.} The process $\nu_\pm\nu_\pm\to hh$ also ceases to
grow with energy because the last diagram in Fig.~\ref{nunuHH}, which was
responsible for the term that grows with energy, is eliminated and replaced by
a diagram analogous to the first diagram of Fig.~\ref{ffVVtriplet} (with the
$V_L$ replaced by $h$).  Thus the scale of Majorana-neutrino mass generation
is the mass of these Higgs scalars.  This is because the theory above the mass
of these scalars, Eq.~(\ref{htriplet}), is renormalizable.

The Higgs potential of the model is discussed in Appendix C.   The triplet
field has a mass term allowed by the gauge symmetry, ${\cal L} = -M_T^2
\Phi^{i*}\Phi^i$, so it is natural for it to be much heavier than the weak
scale, in which case the Higgs scalars $H^0,H^-,H^{--}$ have masses of
approximately $M_T$. The unique renormalizable term in the potential linear in
the triplet field is ${\cal L} = -M_3
\phi^T\epsilon\sigma^i\phi\Phi^{i*}+H.c.$\footnote{This term is absent in the
Majoron model \cite{Gelmini:1981re,Georgi:1981pg}, in which the CP-odd scalar
is the Goldstone boson of spontaneously-broken lepton number. That model is
ruled out by the measurement of the $Z$ width.} In the limit $M_T \gg v$, the
vacuum-expectation value of the triplet field is $u \approx M_3v^2/M_T^2$,
which is much less than $v$. Since the Majorana neutrino mass is $m_\nu = 2y_M
u$, this model provides a natural explanation of why neutrino masses are
light.  Solving for the mass of the heavy Higgs scalars in terms of the
neutrino mass, one obtains $M_T \approx 2(M_3/M_T)y_Mv^2/m_\nu$. This respects
the upper bound on the scale of Majorana-neutrino mass generation,
Eq.~(\ref{nubound}), since $M_3/M_T \ltap \sqrt{\pi}$ (see Appendix C) and
$y_M \ltap \sqrt{2\pi}$ (the analogue of $y_D \ltap \sqrt{8\pi}$ mentioned in
the previous section). The bound is saturated (within a factor of order unity)
when both $M_3/M_T$ and $y_M$ attain their maximum values.

Below the mass of the heavy Higgs scalars, $M_T$, one integrates out the Higgs
triplet field and obtains the dimension-five interaction of
Eq.~(\ref{nuYukawa}), with $c/M = 2M_3y_M/M_T^2$.  Since we associate $M$, the
scale of Majorana-neutrino mass generation, with $M_T$, we are left with
$c=2M_3y_M/M_T$.

The study of these two models leads us to the following definition: {\em The
scale of fermion mass generation is the minimum energy at which the fermion
mass is generated by a renormalizable interaction.}\footnote{This
renormalizable interaction may be supplemented by interactions of dimension
greater than four that also contribute to the fermion mass.}  In the standard
model the fermion mass is generated by a renormalizable interaction at all
energies (above the Higgs mass), so there is no scale of fermion mass
generation.\footnote{Based on this definition, one could argue that the Higgs
mass is the scale of fermion mass generation in the standard model.  As
discussed in Sections~\ref{EWSB} and \ref{standardmodel}, we regard the Higgs
mass as the scale of electroweak symmetry breaking, but not the scale of
fermion mass generation.}

\section{Fermions in non-standard representations}\label{nonstandard}

With the fermion content of the standard model, the only fermions that do not
acquire their mass from a renormalizable interaction with the Higgs field are
Majorana neutrinos, Eq.~(\ref{nuYukawa}).  In this section we extend the
fermion content of the standard model to include fermions in nonstandard
representations of $SU(2)_L\times U(1)_Y$, such that they acquire Dirac masses
from nonrenormalizable interactions.  This will demonstrate that the results
obtained for Majorana neutrinos in the previous section are not peculiar to the
Majorana nature of the fermions.  Furthermore, by choosing the fermion
representation appropriately, we will be able to construct interactions of
arbitrary dimension to generate the fermion mass.  This will allow us to study
the consequences of unitarity in a more general setting.

Consider adding an $SU(2)_L$-triplet, $Y=-1$ fermion field $F_L^{\alpha\beta}$
and an $SU(2)_L$-singlet, $Y=-2$ fermion field $f_R^{--}$ to the standard
model. As it stands, this model has gauge and gravitational anomalies;
however, it is possible to embed this model in an anomaly-free model, as
demonstrated explicitly in Appendix D.  The lowest-dimension interaction that
couples these fermions to the Higgs field ($Y=1/2$) is
\begin{equation}
{\cal L}=-\frac{c}{M}{\overline F_L}^{\alpha\beta}\phi^\alpha\phi^\beta
f_R^{--} +H.c.\;,\label{FYukawa}
\end{equation}
which is the analogue of Eq.~(\ref{Yukawa}), but is of dimension five, like
Eq.~(\ref{nuYukawa}). The $SU(2)_L$-triplet field can be represented by a
symmetric two-index tensor in $SU(2)_L$ space,
\begin{equation}
F_L^{\alpha\beta} \equiv \left(\begin{array}{cc} f_L^0
& \frac{1}{\sqrt 2}f_L^- \\
\frac{1}{\sqrt 2}f_L^- & f_L^{--}\end{array}\right)\;.
\end{equation}
This Lagrangian gives rise to a Dirac mass $m_f=cv^2/2M$ for the field $f^{--}$
when the neutral component of the Higgs field acquires a vacuum-expectation
value $\langle\phi^0\rangle=v/\sqrt 2$.\footnote{One may generate Dirac masses
for the other fields in $F_L^{\alpha\beta}$ by introducing the additional
$SU(2)_L$-singlet fields $f_R^-$ ($Y=-1$)and $f_R^0$ ($Y=0$) and constructing
the analogues of Eq.~(\ref{FYukawa}), making use of the $Y=-1/2$ field
$\epsilon\phi^*$.}

The Feynman diagrams for the amplitude $f^{--}_\pm f^{++}_\pm\to V_LV_L$ are
similar to those in Figs.~(\ref{ffVV}) and (\ref{ffHiggs}).  However, the
$s$-channel Higgs diagram of Fig.~(\ref{ffHiggs}) does not cancel the term
that grows with energy, in contrast to the case of standard-model Dirac
fermions. Thus the situation is analogous to the case of Majorana neutrinos
discussed in the previous section.  This demonstrates that the results
obtained there were not peculiar to the Majorana nature of the fermions, but
instead stem from the fact that the fermion mass is generated by a
dimension-five interaction.

As in the previous section, one can use the Goldstone-boson equivalence theorem
to calculate the high-energy behavior of the amplitude for $f^{--}_\pm
f^{++}_\pm\to V_LV_L$ (as well as the amplitude with one $V_L$ replaced by
$h$). The terms that grow with energy all come from the dimension-five
interaction, Eq.~(\ref{FYukawa}).  This interaction yields a Feynman diagram
similar to the last diagram in Fig.~\ref{nunuHH}.  The resulting amplitude is
proportional to $c\sqrt s/M\approx m_f\sqrt s/v^2$, as in the case of Majorana
neutrinos. The strongest upper bound on the scale of fermion mass generation
comes from applying the inelastic unitarity condition $|a_0|\le 1/2$ to the
amplitude\footnote{The same bound may also be obtained by considering the
amplitude
\begin{eqnarray}
a_0\left(\frac{1}{\sqrt 2}(f^{--}_+f^{++}_+ + f^{--}_-f^{++}_-) \to
Z_Lh\right) \sim -\frac{m_f \sqrt s}{4\pi\sqrt 2 v^2}\;,\nonumber
\end{eqnarray}
which is the analogue of the equation in footnote 8.}
\begin{equation}
a_0\left(\frac{1}{\sqrt 2}(f^{--}_+f^{++}_+ - f^{--}_-f^{++}_-) \to
\frac{1}{2}(Z_LZ_L+hh)\right) \sim -\frac{m_f \sqrt s}{4\pi\sqrt 2 v^2}\;,
\end{equation}
which is the analogue of Eq.~(\ref{a0neutrino}). This yields the upper bound on
the scale of Dirac-fermion mass generation
\begin{equation}
\Lambda_5 \equiv \frac{4\pi v^2}{\sqrt 2m_f}\;,\label{lambda5}
\end{equation}
where the subscript indicates that the Dirac fermion mass is generated by a
dimension-five interaction, Eq.~(\ref{FYukawa}). This is the analogue of
Eq.~(\ref{nubound}).

One can generalize this analysis to an interaction of arbitrary dimension as
follows. Consider the standard model with the addition of an $SU(2)_L$
$(n+1)$-plet, $F_L^{\alpha\cdots\beta}$, with $n$ totally symmetric indices.
Also add an $SU(2)_L$-singlet field $f_R^Q$ of hypercharge (and electric
charge) $Q$.\footnote{The hypercharge of the field $F_L^{\alpha\cdots\beta}$ is
$Y=Q+n/2$.  This model has gauge and gravitational anomalies, but it can be
embedded in an anomaly-free model for some value of $Q$, as we show explicitly
for the $n=2$ case in Appendix D.} The lowest-dimension interaction that
generates a Dirac mass is the dimension-$d$ interaction
\begin{equation}
{\cal L}=-\frac{c}{M^{d-4}}{\overline
F_L}^{\alpha\cdots\beta}\phi^\alpha\cdots\phi^\beta
f_R^Q+H.c.\;,\label{FdYukawa}
\end{equation}
where $d=n+3$.  The fields $F_L^{2\cdots 2}\equiv f_L^Q$ and $f_R^Q$ form a
Dirac mass term of mass $m_f=c(v/\sqrt 2)^{d-3}/M^{d-4}$ when the neutral
component of the Higgs field acquires a vacuum-expect\-ation value
$\langle\phi^0\rangle=v/\sqrt 2$.

\begin{figure}[t]
\begin{center}
\vspace*{0cm} \hspace*{0cm} \epsfxsize=4cm \epsfbox{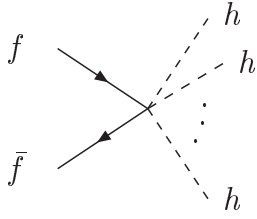} \vspace*{0cm}
\caption{Feynman diagram for the interaction of a fermion with $n$ Higgs
bosons.} \label{ffhh}
\end{center}
\end{figure}

Applying the unitarity condition $|a_0|\le 1/2$ to the amplitude for
$f^Q_\pm\bar f^Q_\pm \to V_LV_L$ (most easily calculated using the
Goldstone-boson equivalence theorem) again yields an upper bound on the scale
of fermion mass generation that is proportional to $v^2/m_f$, like
Eq.~(\ref{lambda5}). However, the strongest upper bound on the scale of
fermion mass generation comes not from this process, but instead from
$f^Q_\pm\bar f^Q_\pm\to h\cdots h$, with $n=d-3$ Higgs bosons in the final
state. The relevant Feynman diagram, shown in Fig.~\ref{ffhh}, is generated
from the dimension-$d$ interaction of Eq.~(\ref{FdYukawa}).  The
unitarity condition on this inelastic amplitude is given in
Eq.~(\ref{unitarity}), where $\sigma_{inel}(2\to n)$ is the total cross section
for $f^Q_\pm\bar f^Q_\pm\to h\cdots h$.  The strongest bound on the scale of
fermion mass generation is obtained by considering the initial state
$(f^Q_+\bar f^Q_+ \mp f^Q_-\bar f^Q_-)/\sqrt 2$, and summing over the cross
sections obtained by replacing an even (upper sign) or odd (lower sign) number of
$h$'s in the final state with $Z_L$'s (or, via the equivalence theorem,
$\chi$'s). Hence, for a Dirac fermion whose mass is
generated via the dimension-$d$ interaction of Eq.~(\ref{FdYukawa}),
the upper bound on the scale of fermion mass generation is
\begin{equation}
\Lambda_d \equiv 4\pi\kappa_d\left(\frac{v^{d-3}}{m_f}\right)^{1/(d-4)}\;,
\label{lambdad}
\end{equation}
where $\kappa_d$, given in Eq.~(\ref{kappad}), is a number of order unity. We
derive this result in Appendix E. The results for $\Lambda_d$ are listed in
Table~\ref{dim} for a few values of $d$.

The scale $M$ has the natural interpretation as the energy at which the
effective field theory involving the dimension-$d$ interaction of
Eq.~(\ref{FdYukawa}) is subsumed by a deeper theory. For example, $M$
corresponds to the mass of the heavy neutrino in the see-saw model discussed
in Section~\ref{neutrino}. Since the fermion acquires a mass $m_f=c(v/\sqrt
2)^{d-3}/M^{d-4}$ from the dimension-$d$ interaction, the scale $M$ is related
to the fermion mass by
\begin{equation}
M=\left(\frac{c}{m_f}\left(\frac{v}{\sqrt 2}\right)^{d-3}\right)^{1/(d-4)}\;.
\end{equation}
Thus $M$ respects the upper bound on the scale of fermion mass generation,
Eq.~(\ref{lambdad}), provided that $c\le \sqrt 2(4\pi\sqrt 2\kappa_d)^{d-4}$.
This condition corresponds to the convergence of the energy expansion, based
on the interaction of Eq.~(\ref{FdYukawa}), for $E \ltap M$.

\begin{table}[t]
\caption{Upper bound on the scale of fermion mass generation, $\Lambda_d$, for
Dirac fermions that acquire a mass via the dimension-$d$ interaction of
Eq.~(\ref{FdYukawa}).}
\begin{center}
\begin{tabular}{l|c}
\hline\hline $d$ & $\Lambda_d$ \\ \hline
4 & $\infty$ \\
5 & $\frac{4\pi}{\sqrt 2} \frac{v^2}{m_f}$ \\
6 & $\frac{4\pi}{6^{1/4}}\left(\frac{v^3}{m_f}\right)^{1/2}$ \\
$\vdots$ & $\vdots$ \\
$d$ & $4\pi\kappa_d\left(\frac{v^{d-3}}{m_f}\right)^{1/(d-4)}$ \\[.3cm]
\hline \hline
\end{tabular}
\end{center}
\label{dim}
\end{table}

\section{Higher-dimension interactions}\label{higherdimension}

In the standard model, Dirac fermions acquire mass via a dimension-four
interaction with the Higgs field, Eq.~(\ref{Yukawa}).  As we argued in Section
\ref{standardmodel}, there is no scale of fermion mass generation in the
standard model.  However, it is likely that the standard model is supplemented
by higher-dimensions interactions, whose presence has not yet been revealed to
us due to the insufficient energy and/or accuracy of our experiments.  In this
section we consider the implications of higher-dimension interactions on the
scale of Dirac-fermion mass generation in the standard model.  Our discussion
applies to all models that reduce to the standard model when the mass of the
physics beyond the standard model is taken to infinity (decoupling).

The lowest-dimension interaction available to supplement the standard model is
of dimension five.  With the usual fermion content (no $\nu_R$ field), there
is only one such interaction, which we already encountered in
Eq.~(\ref{nuYukawa}). This interaction gives rise to a Majorana mass for the
neutrino, but no other fermion masses. Thus we must consider interactions of
at least dimension six in the case of Dirac fermions.

In contrast with interactions of dimension five, there are a large number of
interactions of dimension six available with the field content of the standard
model \cite{Buchmuller:1986jz}. However, there is only one that contributes to
fermion masses, given by
\begin{equation}
{\cal L} = -\frac{c}{M^2}{\overline F_L} \phi f_R\phi^\dagger\phi+H.c.\;,
\label{dimsix}
\end{equation}
which was already considered by Golden \cite{Golden:1994pj}.  This
interaction, in concert with the usual dimension-four interaction of
Eq.~(\ref{Yukawa}), yields a Dirac fermion mass, when the neutral component of
the Higgs field acquires a vacuum-expectation value
$\langle\phi^0\rangle=v/\sqrt 2$, of
\begin{equation}
m_f = y_f \frac{v}{\sqrt 2} + \frac{c}{M^2}\left(\frac{v}{\sqrt 2}\right)^3\;.
\label{dim6} \end{equation}
This interaction also affects the coupling of the
Higgs boson to the fermion, thereby affecting the contribution of the diagram
in Fig.~\ref{ffHiggs} to $f\bar f \to V_LV_L$.  The resulting
zeroth-partial-wave amplitude grows with energy like
\begin{equation}
a_0\left(f_\pm \bar f_\pm \to V_LV_L\right) \approx \frac{c}{M^2}v\sqrt s\;,
\end{equation}
which exceeds the unitarity bound at an energy of order
\begin{equation}
\Lambda \approx \frac{M^2}{cv} \approx \frac{v^2}{m_f-y_fv/\sqrt 2}\;,
\label{Lambdadim6}\end{equation} where we have used Eq.~(\ref{dim6}). This is
an upper bound on the scale of new physics.

In the standard model, where $m_f = y_fv/\sqrt 2$, the upper bound on the scale
of new physics implied by Eq.~(\ref{Lambdadim6}) is infinity.  For
Eq.~(\ref{Lambdadim6}) to imply a scale of new physics, one would need to know
not only that the fermion has a mass $m_f$, but also that the dimension-four
Yukawa coupling of the fermion, Eq.~(\ref{Yukawa}), differs from the
standard-model value $y_f=\sqrt 2m_f/v$.\footnote{This could be inferred by
measuring the coupling of the Higgs boson to the fermion and equating it to
$\sqrt 2 y_f - 3m_f/v$. Only if $y_f=\sqrt 2m_f/v$ will this coupling acquire
the standard-model value $-m_f/v$.} Let us imagine that this Yukawa coupling
were very small, such that the fermion acquires its mass dominantly from the
dimension-six interaction in Eq.~(\ref{dimsix}).  The upper bound on the scale
of new physics implied by Eq.~(\ref{Lambdadim6}) is then proportional to
$v^2/m_f$.

However, as we saw in the previous section, when a fermion acquires a mass via
a dimension-six interaction, a stronger upper bound can be obtained by
considering the unitarity of the process $f_\pm\bar f_\pm \to V_LV_LV_L$. One
finds
\begin{equation}
a_0\left(f_\pm \bar f_\pm \to V_LV_LV_L\right) \approx \frac{c}{M^2}\sqrt s\;,
\end{equation}
which exceeds the unitarity bound of Eq.~(\ref{unitarity}) at an energy of
order
\begin{equation}
\Lambda^2 \approx \frac{M^2}{c} \approx \frac{v^3}{m_f-y_fv/\sqrt 2}\;,
\label{Lambdadim6prime}
\end{equation}
where we have used Eq.~(\ref{dim6}). If we imagine that the Yukawa coupling
were very small, such that the fermion acquires its mass dominantly from the
dimension-six interaction in Eq.~(\ref{dimsix}), then the upper bound on the
scale of new physics is proportional to $(v^3/m_f)^{1/2}$.

In general, both the dimension-four interaction of Eq.~(\ref{Yukawa}) and the
dimension-six interaction of Eq.~(\ref{dimsix}) contribute to the fermion mass.
In keeping with our definition of the scale of fermion mass generation
presented at the end of Section~\ref{neutrino}, we regard
Eq.~(\ref{Lambdadim6prime}) as an upper bound on the scale of new physics, not
an upper bound on the scale of fermion mass generation.  Since the fermion
mass is generated in part by a renormalizable interaction at all energies
(above the Higgs mass), there is no scale of fermion mass generation, as in
the case of the standard model.

As a specific example of a model with a decoupling limit, consider a model with
two Higgs-doublet, $Y=1$ fields, with a discrete symmetry $\phi_1 \to -\phi_1$
such that only $\phi_2$ couples to a given fermion. The most general
scalar potential for this model may be written
as~\cite{Haber:1995be}\footnote{We impose CP-symmetry for simplicity.
This does not affect the generality of our arguments.}
\begin{eqnarray}
V(\phi_1,\phi_2)&=&m_{11}^2\phi_1^\dagger\phi_1 + m_{22}^2\phi_2^\dagger\phi_2
-m_{12}^2[\phi_1^\dagger\phi_2+\phi_2^\dagger\phi_1]
+\frac{1}{2}\lambda_1(\phi_1^\dagger\phi_1)^2
+\frac{1}{2}\lambda_2(\phi_2^\dagger\phi_2)^2 \nonumber \\
&+&\lambda_3(\phi_1^\dagger\phi_1)(\phi_2^\dagger\phi_2)
+\lambda_4(\phi_1^\dagger\phi_2)(\phi_2^\dagger\phi_1)
+\frac{1}{2}\lambda_5[(\phi_1^\dagger\phi_2)^2+(\phi_2^\dagger\phi_1)^2]\;,
\label{potential}
\end{eqnarray}
where the $\lambda_i$'s are real, and where the discrete symmetry is softly
broken by the term proportional to $m_{12}^2$.  The coupling of a fermion $f$
to the Higgs field $\phi_2$ is given by a dimension-four Yukawa interaction
\begin{equation}
{\cal L} =-y_f{\overline F_L} \phi_2 f_R+H.c.\;, \label{Yukawa2}\end{equation}
where $F_L$ is an $SU(2)_L$-doublet fermion field whose lower component is
$f_L$.

We study the decoupling limit in a simple way, by integrating out one of the
Higgs-doublet fields. A convenient way to accomplish this is to first make a
rotation in Higgs-doublet-field space such that the mass matrix is diagonal.
Thus we define fields $\Phi,\phi$, given by
\begin{equation}
\left(\begin{array}{c} \Phi \\ \phi \end{array}\right)=\left(\begin{array}{rr}
\cos\alpha
&  \sin\alpha \\
-\sin\alpha & \cos\alpha \end{array}\right)\left(\begin{array}{c} \phi_1 \\
\phi_2 \end{array}\right)\;,
\end{equation}
where the angle $\alpha$ is chosen to eliminate the off-diagonal term in the
mass matrix, proportional to $m_{12}^2$.\footnote{The angle $\alpha$ is
standard notation in two-Higgs-doublet models \cite{Haber:1995be}.} The
resulting scalar potential is
\begin{equation}
V(\phi,\Phi)=-\mu^2\phi^\dagger\phi+M^2\Phi^\dagger\Phi+\cdots
+\tilde\lambda_6[(\phi^\dagger\phi)(\phi^\dagger\Phi) + H.c.]\;,
\end{equation}
where we have suppressed all quartic interactions except a term, linear in
$\Phi$, which is induced by the rotation in Higgs-field space.  This is the
unique term linear in $\Phi$; its coefficient $\tilde\lambda_6$ is a linear
combination of the $\lambda_i$'s in Eq.~(\ref{potential}).\footnote{$
\tilde{\lambda}_6= \frac{1}{2} \sin 2\alpha
  \left( \lambda_3 + \lambda_4 + \lambda_5 +
    \cos^2 \alpha \,\left( \lambda_2 -
       2\,\left( \lambda_3 + \lambda_4 +
          \lambda_5 \right)  \right)  -
    \lambda_1 \,\sin^2 \alpha \right)\;.$}
We now consider the decoupling limit $M^2 \gg \mu^2$ and integrate out the
Higgs field $\Phi$. In so doing the Yukawa interaction of Eq.~(\ref{Yukawa2})
becomes, for energies less than $M$,
\begin{eqnarray}
{\cal L}&=&-y_f\cos\alpha{\overline F_L} \phi f_R
- y_f\sin\alpha{\overline F_L}\Phi f_R+ H.c. \nonumber \\
&=& -y_f^\prime{\overline F_L} \phi f_R - \frac{c}{M^2}{\overline F_L}\phi f_R
\phi^\dagger\phi + H.c.\;, \label{LE2HDM}
\end{eqnarray}
where $y_f^\prime=y_f\cos\alpha$ and $c=-y_f\tilde\lambda_6\sin\alpha$.  This
interaction is exactly of the form of the standard model plus the dimension-six
term of Eq.~(\ref{dimsix}), where $M$ is identified with the mass of the heavy
Higgs field.

In Ref.~\cite{Jager:1998va}, the decoupling limit of a
two-Higgs-doublet model was studied in an attempt to find a model in which the
scale of fermion mass generation saturates the Appelquist-Chanowitz bound,
$\Lambda_f \approx v^2/m_f$. The mass of the heavy neutral Higgs scalar was
identified as the scale of fermion mass generation. We instead consider it to
be a scale of new physics; there is no scale of fermion mass generation since
the fermion mass arises in part from a renormalizable interaction.  This
attempt to saturate the Appelquist-Chanowitz bound with the mass of the heavy
neutral Higgs scalar failed, and instead Ref.~\cite{Jager:1998va} identified
the upper bound on the mass of this particle to be proportional to
$(v^3/m_f)^{1/2}$, as one would expect if the fermion mass arose from a
dimension-six interaction (see Table~\ref{dim}).  This occurs because in the
limit studied in Refs.~\cite{Jager:1998va,Chivukula:1998zn} one obtains
$\alpha \to -\pi/2$, in which case $y_f^\prime \to 0$ in Eq.~(\ref{LE2HDM}).
The fermion mass is therefore generated by the dimension-six interaction of
Eq.~(\ref{LE2HDM}). Thus we reproduce the results of
Refs.~\cite{Jager:1998va,Chivukula:1998zn} in a much simpler way.

It was also shown in Refs.~\cite{Jager:1998va,Chivukula:1998zn} that the only
limit in which the mass of the heavy neutral Higgs scalar $H$ can saturate the
Appelquist-Chanowitz bound, $\Lambda_f\approx v^2/m_f$, is if some of the
quartic couplings are taken to grow with the heavy Higgs mass
(nondecoupling).  We show in Appendix F that the two models studied in
Refs.~\cite{Jager:1998va} and \cite{Chivukula:1998zn} are not the same,
although they both involve allowing one or more quartic couplings to grow with
the heavy Higgs mass.  However, these models are unphysical since the quartic
couplings cannot exceed ${\cal O}(4\pi)$.

\section{Models without a Higgs field}\label{strong}

The standard model (and extensions thereof) is the unique theory in which the
longitudinal weak vector bosons can be treated as weakly-coupled degrees of
freedom at energies above the scale of electroweak symmetry breaking.  In this
section we discuss the scale of fermion mass generation in models without a
Higgs field. We will see that the upper bound on the scale of fermion mass
generation depends on the dimensionality of the interaction responsible for
generating the fermion mass.

Above energies above $\Lambda_{EWSB}\equiv \sqrt{8\pi}v$, the longitudinal weak
vector bosons cannot generally be treated as weakly-coupled degrees of
freedom.  As discussed in Section \ref{standard}, at high energies one may
think of the longitudinal weak vector bosons as Goldstone bosons via the
Goldstone-boson equivalence theorem.  The situation is analogous to QCD, where
the pions are the Goldstone bosons of broken chiral symmetry. Consider the
process $e^+e^-\to \pi^+\pi^-$.  At energies less than the scale of chiral
symmetry breaking, $\Lambda_{\chi SB} \approx 1$ GeV, one may
treat the pions as point particles, using the effective chiral Lagrangian.
However, above the scale of chiral symmetry breaking, it is invalid to treat
the pions as point particles.\footnote{If one were to do so, one would
conclude that the cross section for $e^+e^-\to \pi^+\pi^-$ falls off like
$1/s$ at high energies.  In fact, this cross section falls much more rapidly
with $s$, due the structure of the pion, which yields a form factor for the
photon-pion interaction.  The pion form factor, $F_\pi(s)$, is believed to fall
off like $1/s$ at large $s$ \cite{Lepage:1980fj}.  This yields a cross section
that falls off like $1/s^3$.} In the same way, the electroweak model ceases to
be a useful description of longitudinal weak vector bosons at energies above
the scale of electroweak symmetry breaking if there is no Higgs field.

Consider fermion mass generation in a theory in which electroweak symmetry
breaking is described by technicolor \cite{Weinberg:1979bn,Susskind:1979ms}.
Since the longitudinal weak vector bosons are not weakly coupled above
$\Lambda_{EWSB}$, one cannot calculate amplitudes involving external
longitudinal weak vector bosons perturbatively. However, one may still discuss
the scale of fermion mass generation. At the weak scale the lowest-dimension
interaction that generates a fermion mass is a dimension-six interaction
between technifermions and ordinary fermions, which yields a fermion mass when
the technifermions condense.  If the coefficient of this dimension-six
interaction is $c/M^2$, one obtains
\begin{equation}
m_f \approx c\frac{\langle \overline TT \rangle}{M^2}\;, \label{ETC}
\end{equation}
where $\langle \overline TT \rangle$ is the technifermion condensate.  In
extended technicolor (ETC), this dimension-six interaction is the low-energy
approximation to the interaction of fermions and technifermions via the
exchange of extended-technicolor gauge bosons of mass $M_{ETC}$
\cite{Eichten:1980ah,Dimopoulos:1979es}.  Since the theory above $M_{ETC}$ is
renormalizable, the scale of fermion mass generation is $M_{ETC}$.  We identify
$M$ with $M_{ETC}$, and $c$ with $g_{ETC}^2$, the square of the ETC gauge
coupling. Thus one obtains from Eq.~(\ref{ETC})
\begin{equation}
M_{ETC} \approx \left(g_{ETC}^2\frac{\langle \overline TT
\rangle}{m_f}\right)^{1/2}\;. \label{METC}
\end{equation}
In a QCD-like model $\langle \overline TT \rangle \approx v^3$.  Thus
Eq.~(\ref{METC}) is the analogue of the upper bound on the scale of fermion
mass generation obtained in the model of Section \ref{nonstandard} in which a
fermion acquires a mass from a dimension-six interaction, $\Lambda_6 \approx
(v^3/m_f)^{1/2}$ (see Table~\ref{dim}).

The scale of fermion mass generation, $M$, can be increased
for a fixed value of $m_f$ if the technifermion condensate, evaluated at $M$,
is enhanced. Such is the case in walking technicolor
\cite{Holdom:1981rm,Holdom:1985sk,Yamawaki:1986zg,Appelquist:1986an}. This may
also be described in terms of the dimensionality of the operator responsible
for generating the fermion mass. The composite operator $\overline TT$ has a
large anomalous dimension, $\gamma_m>1$, which is assumed to be constant over
the range of energies $v \ltap E \ltap M$.  Thus the four-fermion operator
responsible for generating the fermion mass has scaling dimension $6-\gamma_m$
over this range. The fermion mass is given by
\begin{equation}
m_f \approx c\frac{\langle \overline TT
\rangle}{M^2}\left(\frac{M}{v}\right)^{\gamma_m}\;, \label{WETC}
\end{equation}
so the scale of fermion mass generation is related to the fermion mass by
\begin{equation}
M \approx \left(c\frac{v^{3-\gamma_m}}{m_f}\right)^{1/(2-\gamma_m)}\;,
\label{WMETC}
\end{equation}
where we have used $\langle \overline TT \rangle \approx v^3$.\footnote{This is
the value of the condensate evaluated at the weak scale.}  This is the
analogue of $\Lambda_d\approx (v^{d-3}/m_f)^{1/(d-4)}$, Eq.~(\ref{lambdad}),
for an interaction of scaling dimension $d=6-\gamma_m$.

A particularly interesting case of Eq.~(\ref{WETC}) occurs when the physics at
$M$ is fine tuned such that $\gamma_m =
2$~\cite{Appelquist:1989as,Takeuchi:1989qa,Miransky:1989gk}. In this case, the
enhancement of the technifermion condensate exactly cancels the $1/M^2$
suppression of the four-fermion operator responsible for generating the
fermion mass, leading to $m_f = {\cal O}(v)$, independently of the value of
$M$.  Hence, there is no upper bound on the scale of fermion mass generation,
as also follows from Eq.~(\ref{WMETC}). The scaling dimension of the composite
operator $\overline TT$ becomes $3-\gamma_m = 1$ in this case, the same as
that of a weakly-coupled scalar field.  It is natural to associate this
fine-tuned limit with the emergence of a light, composite scalar that acquires
a small vacuum-expectation value $v\ll M$ and that has renormalizable Yukawa
couplings (unsuppressed by $M$) to standard-model
fermions~\cite{Chivukula:1990bc}. At energies less than $M$, this composite
scalar behaves like a Higgs boson, and the resulting theory reduces to the
standard model when $M$ is taken to infinity (decoupling). Accordingly, the
considerations of Sections~\ref{standardmodel} and \ref{higherdimension} apply,
where we concluded that there is no upper bound on the scale of fermion mass
generation, in agreement with the above argument.

\section{Conclusions}\label{conclusions}

In this paper we studied the scale of fermion mass generation.  We critically
re-examined an upper bound on this scale, due to Appelquist and Chanowitz
\cite{Appelquist:1987cf}, obtained by considering the process $f_\pm\bar
f_\pm\to V_LV_L$, where the subscripts on the fermions indicate helicity $\pm
1/2$, and $V_L=W_L,Z_L$ denotes longitudinal (helicity zero) weak vector
bosons.  In the absence of the Higgs boson, the amplitude for this process
grows with energy and violates the unitarity bound at an energy of order
$\Lambda_f \approx v^2/m_f$.  We showed that there exists a stronger bound,
proportional to $(v^n/m_f)^{1/(n-1)}\!$, obtained by considering the process
$f_\pm\bar f_\pm\to V_L \cdots V_L$ with $n>2$ particles in the final state.
For large $n$, this bound is arbitrarily close to the upper bound on the scale
of electroweak symmetry breaking, regardless of the fermion mass.  Thus there
is no upper bound on the scale of fermion mass generation.

We further argued that the derivation of this bound is valid only if the
longitudinal weak vector bosons are weakly coupled at energies above the scale
of electroweak symmetry breaking. This requires the existence of a Higgs
doublet, since the standard Higgs model (and extensions thereof that decouple
when the mass of the additional physics is taken to infinity) is the unique
theory in which the longitudinal weak bosons remain weakly coupled at high
energy. Once the Higgs doublet is included in the theory, the upper bound on
the scale of fermion mass generation depends only on the dimensionality of the
operator responsible for generating the fermion mass.  In the standard model,
fermions acquire their mass from a dimension-four interaction with the Higgs
field, which has a dimensionless Yukawa coupling. Thus there is {\em no} scale
of fermion mass generation in the standard model.

Majorana neutrinos acquire their mass from an interaction of dimension five,
with a coefficient with dimensions of an inverse power of mass.  This mass
sets the scale for Majorana-neutrino mass generation.  The amplitude for
$\nu_\pm\nu_\pm\to V_LV_L$ grows with energy despite the inclusion of the
Higgs boson, because the neutrino acquires its mass from a nonrenormalizable
interaction.  Applying the unitarity condition to the amplitude, we derived an
upper bound on the scale of Majorana-neutrino mass generation
\cite{Maltoni:2001iq}
\begin{equation}
\Lambda_{Maj} \equiv \frac{4\pi v^2}{m_\nu}\;. \label{LambdaMaj}
\end{equation}
The upper bounds on $\Lambda_{Maj}$ implied by a variety of neutrino-oscillation
experiments are listed in Table~\ref{massdiffs}.

We considered extending the standard model by adding fermions in nonstandard
representations of $SU(2)_L\times U(1)_Y$ such that they acquire a Dirac mass
from an interaction of dimension $d$.  We showed that the strictest upper
bound on the scale of fermion mass generation is obtained by applying the
unitarity condition to the amplitude for $f_\pm\bar f_\pm\to V_L\cdots V_L$,
with $n=d-3$ particles in the final state.  This upper bound is proportional to
$(v^{d-3}/m_f)^{1/(d-4)}$. For a fermion that acquires mass via the
dimension-$d$ interaction of Eq.~(\ref{FdYukawa}), the upper bound on the
scale of fermion mass generation is listed in Table~\ref{dim}.

For a fermion that acquires its mass via an interaction of dimension four, the
amplitude for $f_\pm\bar f_\pm\to V_LV_L$ ceases to grow with energy above the
Higgs mass.  This reflects the fact that the Higgs mass is the scale of
electroweak symmetry breaking and that the fermion mass is generated via a
renormalizable interaction.  However, the Higgs mass is not the scale of
fermion mass generation, as evidenced by the fact that there is no
cancellation of the term that grows with energy for fermions that acquire their
mass via an interaction of dimension $d>4$.

We defined the scale of fermion mass generation as the minimum energy at which
the fermion mass is generated by a renormalizable interaction.  In the
standard model the fermion mass is generated by a renormalizable interaction
at all energies (above the Higgs mass), so there is no scale of fermion mass
generation.

We also considered extending the standard model by maintaining the same
particle content but adding higher-dimension interactions.  For fermions other
than Majorana neutrinos, the lowest-dimension interaction one can add is of
dimension six.  There is only one dimension-six interaction that affects the
fermion mass.  To learn of the presence of this interaction requires knowledge
not only of the fermion mass, but of its interaction with the Higgs boson.
This will be a goal of future experiments once the Higgs boson is discovered.
We showed that a two-Higgs-doublet model generates this dimension-six
interaction when one of the Higgs doublets is taken to be heavy and is
integrated out.

Finally, we considered models without a Higgs field.  The process $f_\pm\bar
f_\pm\to V_LV_L$ cannot be used to derive an upper bound on the scale of
fermion mass generation because the longitudinal weak vector bosons are not
weakly coupled above the scale of electroweak symmetry breaking. Nevertheless,
one can discuss the scale of fermion mass generation in specific models.  We
showed that the relation between the fermion mass and the scale of fermion
mass generation depends on the dimensionality of the interaction responsible
for generating the fermion mass.

The most important conclusion of this study is that there is no upper bound on
the scale of (Dirac) fermion mass generation in the standard model.  This is
disappointing, because an upper bound on this scale would provide a target for
future accelerators, in the same way that the upper bound on the scale of
electroweak symmetry breaking, $\Lambda_{EWSB} \equiv \sqrt{8\pi} v \approx 1$
TeV, provides a target for the CERN Large Hadron Collider.  This does not
preclude the possibility that new physics lies at accessible energies; it only
says that (Dirac) fermion masses do not imply a scale of new physics. In
contrast, there is an upper bound on the scale of Majorana-neutrino mass
generation, Eq.~(\ref{LambdaMaj}), and although this upper bound is beyond the
reach of future accelerators, the fact that the upper bounds on $\Lambda_{Maj}$
lie near the grand-unification scale (see Table~\ref{massdiffs}) bolsters our
belief in the relevance of grand unification for physics beyond the standard
model.

\section*{Acknowledgments}

We are grateful for conversations with T.~Appelquist, T.~Han, and R.~Leigh.
This work was supported in part by the U.~S.~Department of Energy under
contract No.~DOE~DE-FG02-91ER40677. We gratefully acknowledge the support of
GAANN, under Grant No.~DE-P200A980724, from the U.~S.~Department of Education
for J.~M.~N.

\section*{Appendix A}

We derive the upper bound on the inelastic $2\to n$ scattering cross section,
Eq.~(\ref{unitarity}), from the unitarity of the $S$ matrix, $S^\dagger S=1$.
Writing $S=1+iT$, one obtains
\begin{equation}
T^\dagger T=2\,{\rm Im}\,T\;.
\end{equation}
Take the matrix element of this equation between identical initial and final
two-body states.  Insert a complete set of intermediate states into the
left-hand side of this equation, separating out explicitly the intermediate
state which is identical to the initial and final states, to get
\begin{equation}
\int dPS_2 |T_{el}(2\to 2)|^2 + \sum_n\int dPS_n |T_{inel}(2\to n)|^2 = 2\,{\rm
Im}\,T_{el}(2\to 2)\;,
\end{equation}
where $dPS_n$ indicates $n$-body phase space and the sum is over all inelastic
intermediate states. Define the $J^{\rm th}$ partial-wave $2\to 2$ elastic
amplitude
\begin{equation}
a_J=\frac{1}{32\pi}\int^1_{-1} dz\,P_J(z) T_{el}(2\to 2)\;,
\end{equation}
where $z$ is the cosine of the scattering angle, to get
\begin{equation}
\sum_J |a_J|^2 + \frac{1}{32\pi}\sum_n\int dPS_n |T_{inel}(2\to n)|^2 = \sum_J
{\rm Im}\,a_J\;.
\end{equation}
Using $|a_J|^2=({\rm Re}\,a_J)^2+({\rm Im}\,a_J)^2$ yields
\begin{equation}
\sum_J ({\rm Re}\,a_J)^2+\frac{1}{32\pi}\sum_n\int dPS_n |T_{inel}(2\to n)|^2
=\sum_J {\rm Im}\,a_J(1-{\rm Im}\,a_J)\;.
\end{equation}
If the elastic amplitude is dominated by a single partial wave (J=0 in the
case studied in Section~\ref{standard}), one may remove the summation.  The
right-hand side is then bounded above by $1/4$, yielding
\begin{equation}
\int dPS_n |T_{inel}(2\to n)|^2 \le 8\pi\;, \label{nbound}
\end{equation}
for all $n$.  This implies the desired upper bound,
\begin{equation}
\sigma_{inel}(2\to n) \le \frac{4\pi}{s}\;.\label{unitarity2}
\end{equation}
If there is more than one $n$-body intermediate state, then the bound applies
to the sum of the cross sections for each intermediate state.

\section*{Appendix B}

\begin{figure}[t]
\begin{center}
\vspace*{0cm} \hspace*{0cm} \epsfxsize=13cm \epsfbox{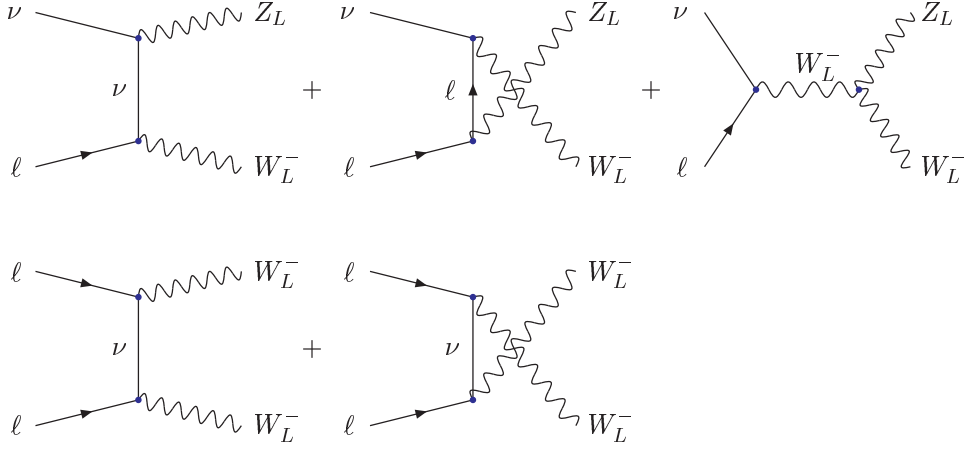} \vspace*{0cm}
\caption{Feynman diagrams that contribute to the amplitude for $\ell\nu\to
W^-_LZ_L$ and $\ell\ell\to W^-_LW^-_L$ in unitary gauge.} \label{nuell}
\end{center}
\end{figure}

We give the high-energy limit of the helicity amplitudes for same-helicity
Majorana-neutrino and charged-lepton scattering into longitudinal weak vector
bosons and Higgs bosons, in a theory in which the Majorana-neutrino mass is
generated by the dimension-five interaction of Eq.~(\ref{nuYukawa}).  The
relevant Feynman diagrams for $\nu\nu$ scattering are shown in
Figs.~\ref{nunuVV}--\ref{nunuHH}; the diagrams for $\ell\nu$ and $\ell\ell$
scattering are given in Fig.~\ref{nuell}.  Our conventions are as follows. We
use a chiral basis for the Dirac matrices and spinors:
\begin{equation}
\gamma^\mu=\left(\begin{array}{cc} 0 &  \sigma^\mu \\
\bar\sigma^\mu & 0 \end{array}\right) \;,
\end{equation}
where $\sigma^\mu = (1,\sigma^i)$, $\bar\sigma^\mu = (1,-\sigma^i)$.  The
spinors for the incoming particles are chosen to be eigenstates of helicity
and read
\begin{eqnarray}
&&u_+(p) = \left(
\begin{array}{cl}
~  \sqrt{E- {\rm p}} & \xi_+ \\
    \sqrt{E+ {\rm p}} & \xi_+ \\
\end{array}
\right)\, \qquad u_-(p) = \left(
\begin{array}{cl}
~  \sqrt{E+ {\rm p}} & \xi_- \\
  \sqrt{E- {\rm p}} & \xi_- \\
\end{array}
\right) \,\\[10pt]
&&v_+(p) = \left(
\begin{array}{cl}
  ~ \sqrt{E+ {\rm p}} & \eta_+ \\
  \!-\sqrt{E- {\rm p}} & \eta_+ \\
\end{array}
\right) \,\qquad v_-(p) = \left(
\begin{array}{cl}
   ~\sqrt{E- {\rm p}} &\eta_- \\
  \!-\sqrt{E+ {\rm p}}  & \eta_- \\
\end{array}
\right)\,
\end{eqnarray}
where $p^\mu=(E, {\rm p} \,\sin\theta\; \cos\phi, {\rm p} \,\sin\theta\;
\sin\phi, {\rm p} \cos\theta)$ and the Pauli spinors $\xi$ and $\eta$ are
defined as follows:
\begin{eqnarray}
&&\xi_+ = \left(
\begin{array}{c}
 \cos \frac{\theta}{2} \\
 e^{i \phi} \sin \frac{\theta}{2}\\
\end{array}
\right) \,,\qquad \xi_- = \left(
\begin{array}{c}
 - e^{-i \phi} \sin \frac{\theta}{2} \\
  \cos \frac{\theta}{2}\\
\end{array}
\right)\,, \qquad \eta_\pm =\pm \xi_\mp\,.
\end{eqnarray}
In the amplitudes listed below, the first fermion has momentum along the
direction $\theta=0,\phi=0$, and the second along the direction
$\theta=\pi,\phi=\pi$.

The zeroth partial-wave amplitudes, in the high-energy limit, are
\begin{eqnarray}
a_0\left(\frac{1}{\sqrt 2}\nu_{i\pm} \nu_{j\pm} \to W^+_LW^-_L\right) &
\sim & 0 \label{WW}\\
a_0\left(\frac{1}{\sqrt 2}\nu_{i\pm} \nu_{j\pm} \to \frac{1}{\sqrt
2}Z_LZ_L\right) &
\sim & \mp \frac{m_{\nu_i} \sqrt s}{16\pi v^2}\delta_{ij} \label{ZZ}\\
a_0\left(\frac{1}{\sqrt 2}\nu_{i\pm} \nu_{j\pm} \to Z_Lh\right) & \sim &
-\frac{m_{\nu_i} \sqrt s}{8\pi\sqrt 2v^2}\delta_{ij}
\label{ZH}\\
a_0\left(\frac{1}{\sqrt 2}\nu_{i\pm} \nu_{j\pm} \to \frac{1}{\sqrt 2}hh\right)
&
\sim & \mp \frac{m_{\nu_i} \sqrt s}{16\pi v^2}\delta_{ij} \label{HH}\\
a_0(\ell_- \nu_{i-} \to W^-_LZ_L) &
\sim & \frac{m_{\nu_i} \sqrt s}{8\pi\sqrt 2v^2}U^*_{\ell i} \label{ZW}\\
a_0(\ell_- \nu_{i-} \to W^-_Lh) &
\sim & - \frac{m_{\nu_i} \sqrt s}{8\pi\sqrt 2v^2}U^*_{\ell i} \label{HW}\\
a_0\left(\frac{1}{\sqrt 2}\ell_- \ell_- \to \frac{1}{\sqrt 2}W^-_LW^-_L\right)
& \sim & \frac{\sqrt s}{8\pi v^2}\sum_{i=1}^3 U_{\ell i}^2 m_{\nu_i}
\;,\label{SS}
\end{eqnarray}
where $v=(\sqrt 2 G_F)^{-1/2}$ is the weak scale, the indices $i,j$ denote the
three neutrino mass eigenstates, the subscripts on the neutrinos and charged
leptons indicate helicity $\pm 1/2$, and the subscript on the partial-wave
amplitudes indicates $J=0$. The unitary matrix $U_{\ell i}$ relates the
neutrino weak and mass eigenstates. Each amplitude grows linearly with energy,
and is proportional to the Majorana-neutrino mass or a linear combination of
masses.

\section*{Appendix C}

In Section \ref{neutrino} we considered a model for Majorana neutrino masses
involving a Higgs doublet, $Y=1/2$ field, $\phi$, and a Higgs triplet, $Y=1$
field, $\Phi^i$.  Here we discuss the scalar potential of this model.

The most general potential is
\cite{Cheng:1980qt,Gelmini:1981re,Mohapatra:1981yp,Georgi:1981pg,Ma:1998dx}
\begin{eqnarray}
V(\phi,\Phi^i)&=&m^2\phi^\dagger\phi+M_T^2\Phi^{i*}\Phi^i
+\lambda_1(\phi^\dagger\phi)^2+\lambda_2(\Phi^{i*}\Phi^i)^2
+2\lambda_3(\phi^\dagger\phi)(\Phi^{i*}\Phi^i) \nonumber \\
&+&\lambda_4(\Phi^i\Phi^i)(\Phi^{j*}\Phi^{j*})
-2i\lambda_5\epsilon^{ijk}\phi^\dagger\sigma^i\phi\Phi^{j*}\Phi^k \nonumber \\
&+&(M_3\phi^T\epsilon\sigma^i\phi\Phi^{i*}+H.c.)\;.
\end{eqnarray}
Minimizing the potential such that the neutral component of the Higgs doublet
acquires a vacuum-expectation value $\langle \phi^0 \rangle=v/\sqrt 2$ and the
neutral component of the Higgs triplet, $\Phi^0=(\Phi^1+i\Phi^2)/\sqrt 2$,
acquires a vacuum-expectation value $\langle \Phi^0 \rangle = u/\sqrt 2$ yields
\begin{equation}
m^2+\lambda_1v^2+(\lambda_3+\lambda_5)u^2-2M_3u=0\;,
\end{equation}
\begin{equation}
M_T^2+\lambda_2u^2+(\lambda_3+\lambda_5)v^2-M_3v^2/u=0\;.
\end{equation}
In the limit that the mass of the Higgs-triplet field, $M_T$, is much greater
than $v$, the equation above implies $u\approx M_3v^2/M_T^2 \ll v$.  Thus the
small value of the vacuum-expectation value of the Higgs triplet field, $u$,
can be understood as a consequence of the large value of the Higgs-triplet
mass, $M_T$ \cite{Ma:1998dx}.

The mass matrix of the scalar fields $\sqrt 2 {\rm Re}\,\phi^0$, $\sqrt 2{\rm
Re}\,\Phi^0$, evaluated at the minimum of the potential, is
\begin{equation}
{\cal M}^2=\left(\begin{array}{cc} 2\lambda_1v^2
& 2(\lambda_3+\lambda_5)uv-2M_3v \\
2(\lambda_3+\lambda_5)uv-2M_3v & 2\lambda_2u^2+M_3v^2/u
\end{array}\right)\;.
\end{equation}
The eigenvalues of this matrix are the masses of the physical scalar bosons,
which must be positive.  Evaluating the determinant of this matrix in the
limit $M_T \gg v \gg u$ gives
\begin{equation}
{\rm det} {\cal M}^2 = 2\lambda_1 v^2M_T^2 - 4 M_3^2v^2 > 0\;.
\end{equation}
This equation, along with the upper bound on the Higgs self coupling,
$\lambda_1 \ltap 2\pi$ \cite{Dashen:1983ts,Luscher:1988gc}, implies the bound
\begin{equation}
\frac{M_3}{M_T} \ltap \sqrt\pi\;,
\end{equation}
which was used in Section \ref{neutrino}.

\section*{Appendix D}

The model presented in Section~\ref{nonstandard} containing an
$SU(2)_L$-triplet, $Y=-1$ fermion field $F_L^{\alpha\beta}$ and an
$SU(2)_L$-singlet, $Y=-2$ fermion field $f_R^{--}$ has gauge and gravitational
anomalies, and is therefore not a consistent theory. However, it is
straightforward to embed this model in a theory with additional fermion fields
such that it is free of all gauge and gravitational anomalies. The fermion
content of this model is given in Table~\ref{model}, with the right-chiral
fermion fields $(F_R^c)^{\alpha\beta}\equiv C\gamma^0F_L^{*\alpha\beta}$ and
$f_R^{--}$ indicated. One can check explicitly that all anomalies cancel
(including the discrete $SU(2)_L$ anomaly \cite{Witten:1982fp}).

The model was constructed as follows.\footnote{See the tables in
Ref.~\cite{Slansky:1981yr}. Our convention for $U(1)$ charges is $-1/2$ of the
convention used in that reference.  In our convention, $Q=T_{3L}+Y$, where $Q$
is electric charge, $Y$ is hypercharge, and $T_{3L}=\pm 1/2$ for $SU(2)_L$
doublets and $1,0,-1$ for $SU(2)_L$ triplets.}  One is seeking a chiral,
anomaly-free $SU(3)\times SU(2)_L\times U(1)_Y$ theory containing an $SU(2)_L$
triplet.  The smallest group with chiral, anomaly-free irreducible
representations is $SO(10)$, and the smallest representation containing an
$SU(2)$ triplet is the 126, which decomposes into the subgroup $SU(4)\times
SU(2)\times SU(2)$ as
\begin{equation}
126=(6,1,1)+(\overline{10},3,1)+(10,1,3)+(15,2,2)\;. \nonumber
\end{equation}
The $(6,1,1)$ and $(15,2,2)$ are real representations, and hence are
automatically anomaly free.  The $10$ and $\overline{10}$ of $SU(4)$ decompose
into the subgroup $SU(3)\times U(1)$ as
\begin{eqnarray}
10&=&1(-1)+3(-1/3)+6(1/3) \nonumber\\
\overline{10}&=&1(1)+\bar 3(1/3)+\bar 6(-1/3) \nonumber
\end{eqnarray}
and the $3$ of $SU(2)$ decomposes into the subgroup $U(1)$ as
\begin{equation}
3=(1)+(0)+(-1)\;. \nonumber
\end{equation}
Consider the decomposition $SO(10)\to SU(4)\times SU(2)\times SU(2) \to
SU(3)\times SU(2)_L\times U(1)_Y$. We identify $SU(2)_L$ with the first
$SU(2)$ and $U(1)_Y$ with the diagonal subgroup of the $U(1)$'s coming from
the decomposition of $SU(4)$ and the second $SU(2)$ (the hypercharge is thus
the sum of the two $U(1)$ charges). This yields the model in Table~\ref{model}.

\begin{table}[t]
\caption{$SU(3)\times SU(2)_L\times U(1)_Y$ representations of an anomaly-free
model containing an $SU(2)_L$-triplet, $Y=1$ fermion field
$(F_R^c)^{\alpha\beta}$ and an $SU(2)_L$-singlet, $Y=-2$ fermion field
$f_R^{--}$.}
\begin{center}
\begin{tabular}{l|ccc}
\hline\hline
& $SU(3)$ & $SU(2)_L$ & $U(1)_Y$ \\ \hline &&&\\[-.4cm]
$(F_R^c)^{\alpha\beta}$&1&3&1\\
$f_R^{--}$&$1$&$1$&$-2$\\
&${\bar 3}$&$3$&$1/3$\\
&${\bar 6}$&$3$&$-1/3$\\
&$1$&$1$&$0$\\
&$3$&$1$&$2/3$\\
&$3$&$1$&$-1/3$\\
&$3$&$1$&$-4/3$\\
&$6$&$1$&$4/3$\\
&$6$&$1$&$1/3$\\
&$6$&$1$&$-2/3$\\
\hline \hline
\end{tabular}
\end{center}
\label{model}
\end{table}

\section*{Appendix E}

We derive the upper bound on the scale of Dirac-fermion mass generation,
Eq.~(\ref{lambdad}), in a model in which the fermion acquires a mass from the
dimension-$d$ interaction of Eq.~(\ref{FdYukawa}).  The bound is obtained by
applying the inelastic unitarity condition, Eq.~(\ref{unitarity})
(Eq.~(\ref{unitarity2}) in Appendix A), to the scattering process $f_\pm^Q\bar
f_\pm^Q \to h\cdots h$ and to the related processes in which some of the $h$'s
are replaced by $Z_L$'s.

Begin with the dimension-$d$ interaction of Eq.~(\ref{FdYukawa}),
\begin{equation}
{\cal L}=-\frac{c}{M^{d-4}}{\overline
F_L}^{\alpha\cdots\beta}\phi^\alpha\cdots\phi^\beta f_R^Q+H.c.\;,
\end{equation}
where there are $n=d-3$ Higgs fields.  Let $\phi = (-is^+,(h+v+i\chi)/\sqrt
2)$, where $s^+,\chi$ are the Goldstone bosons associated with $W^+,Z$.  Using
the Goldstone-boson equivalence theorem, we let $\chi$ represent $Z_L$, and
multiply by a factor of $i$ for each outgoing $Z_L$. The interaction of $m$
neutral Goldstone bosons with $n-m$ Higgs bosons is
\begin{equation}
{\cal L}=-\frac{c}{M^{d-4}}\left(\frac{1}{\sqrt 2}\right)^n {n \choose m} \bar
f^Q \gamma_5^mf^Q (h)^{n-m}(i\chi)^m\;,
\end{equation}
where $f_L^Q\equiv F_L^{2\cdots 2}$.  The fermion acquires a mass
\begin{equation}
m=\frac{c}{M^{d-4}}\left(\frac{v}{\sqrt 2}\right)^n\;,
\end{equation}
so the Feynman rule for the $\bar f^Qf^Q(h)^{n-m}(\chi)^m$ vertex can be
written as [$-i(m/v^n)i^mn!\gamma_5^m$], where we have properly accounted for
the $m$ identical $\chi$'s and the $n-m$ identical $h$'s.

Consider the scattering process
\begin{equation}
2\to n\equiv \frac{1}{\sqrt 2}(f^Q_+\bar f^Q_+ \mp f^Q_-\bar f^Q_-) \to
\frac{1}{\sqrt{(n-m)!}\sqrt{m!}}(h)^{n-m}(\chi)^m\;,
\end{equation}
where the upper (lower) sign corresponds to final states with an even (odd)
number of $\chi$'s. The inelastic unitarity bound, Eq.~(\ref{unitarity})
(Eq.~(\ref{unitarity2}) in Appendix A, or, equivalently, Eq.~(\ref{nbound})),
yields
\begin{equation}
\left(\frac{1}{(n-1)!(n-2)!}\right)\left(\frac{1}{(2\pi)^3}\right)^n(2\pi)^4\frac{\pi}{2}
\left(\frac{\pi}{2}s\right)^{n-2}\frac{1}{(n-m)!m!}\frac{m^2}{v^{2n}}2(n!)^2s\le
8\pi\;,
\end{equation}
where the first five factors are from $n$-body phase space.  Summing over all
$2\to n$ processes with $n-m$ $h$'s and $m$ $\chi$'s (with $m$ either even or
odd), using
\begin{equation}
\sum_{m=0,2,\dots}^n {n \choose m} = \sum_{m=1,3,\dots}^n {n \choose m}
=2^{n-1}\;,
\end{equation}
gives
\begin{equation}
\left(\frac{1}{(n-1)!(n-2)!}\right)\left(\frac{1}{(2\pi)^3}\right)^n(2\pi)^4\frac{\pi}{2}
\left(\frac{\pi}{2}s\right)^{n-2}\frac{m^2}{v^{2n}}2n!2^{n-1}s\le 8\pi\;.
\end{equation}
Defining $\Lambda_d$ as the energy, $\sqrt s$, at which this inequality is
saturated yields Eq.~(\ref{lambdad}),
\begin{equation}
\Lambda_d \equiv 4\pi\kappa_d\left(\frac{v^{d-3}}{m_f}\right)^{1/(d-4)}\;,
\end{equation}
where we have used $n=d-3$ and
\begin{equation}
\kappa_d\equiv\left(\sqrt{\frac{(d-5)!}{2^{d-5}(d-3)}}\right)^{1/(d-4)}\;.
\label{kappad}
\end{equation}

\section*{Appendix F}

In Refs.~\cite{Jager:1998va,Chivukula:1998zn} a two-Higgs-doublet model was
studied in the limit that the mass of the Higgs scalar $H$ is large and one or
more quartic couplings grows with the mass of this Higgs scalar.  The limits
studied in those papers appear to be the same.  Here we show that they are
actually different limits.   Nevertheless, they are both unphysical because
they require a dimensionless coupling to exceed ${\cal O}(4\pi)$.

The Higgs potential used in Ref.~\cite{Jager:1998va} is given in
Eq.~(\ref{potential}).  In Ref.~\cite{Chivukula:1998zn}, a different but
physically equivalent parameterization of the Higgs potential is used
\cite{Georgi:1978xz}:
\begin{eqnarray}
V(\phi_1,\phi_2)&=&\lambda_1^\prime(\phi_1^\dagger\phi_1-v_1^2/2)^2
+\lambda_2^\prime(\phi_2^\dagger\phi_2-v_2^2/2)^2 \nonumber \\
&+&\lambda_3^\prime[(\phi_1^\dagger\phi_1-v_1^2/2)
+(\phi_2^\dagger\phi_2-v_2^2/2)]^2
+\lambda_4^\prime[(\phi_1^\dagger\phi_1)(\phi_2^\dagger\phi_2)
-(\phi_1^\dagger\phi_2)(\phi_2^\dagger\phi_1)] \nonumber \\
&+&\lambda_5^\prime[{\rm Re}\,(\phi_1^\dagger\phi_2)-v_1v_2/2]^2
+\lambda_6^\prime[{\rm Im}\,(\phi_1^\dagger\phi_2)]^2\;.
\label{potentialprime}\end{eqnarray} The coefficients are labeled
$\lambda_i^\prime$ to distinguish them from the coefficients $\lambda_i$ in
Eq.~(\ref{potential}). They are related to the parameters of the Higgs
potential given in Eq.~(\ref{potential}) by
\begin{eqnarray}
\lambda_1&=&2(\lambda_1^\prime+\lambda_3^\prime) \nonumber \\
\lambda_2&=&2(\lambda_2^\prime+\lambda_3^\prime) \nonumber \\
\lambda_3&=&2\lambda_3^\prime+\lambda_4^\prime \nonumber \\
\lambda_4&=&\frac{1}{2}(\lambda_5^\prime+\lambda_6^\prime)-\lambda_4^\prime
\nonumber \\
\lambda_5&=&\frac{1}{2}(\lambda_5^\prime-\lambda_6^\prime) \nonumber \\
m_{11}^2&=&-v_1^2(\lambda_1^\prime+\lambda_3^\prime)-v_2^2\lambda_3^\prime
\nonumber \\
m_{22}^2&=&-v_2^2(\lambda_2^\prime+\lambda_3^\prime)-v_1^2\lambda_3^\prime
\nonumber \\
m_{12}^2&=&\frac{1}{2}v_1v_2\lambda_5^\prime \nonumber
\end{eqnarray}

The limit studied in Ref.~\cite{Chivukula:1998zn} corresponds to taking $m_{H}$
large by letting $\lambda_5^\prime\to\infty$ , since they are approximately
related by
\begin{equation}
m_{H}^2\approx\frac{1}{2} \lambda_5^\prime v^2\;,
\end{equation}
where $v= (\sqrt 2 G_F)^{-1/2}$ is the weak scale.  In terms of the
parameterization of Eq.~(\ref{potential}), used in Ref.~\cite{Jager:1998va},
this limit corresponds to $\lambda_4 = \lambda_5 = m_{12}^2/v_1v_2 \to\infty$,
as is evident from the above relations.  This differs from the limit studied in
Ref.~\cite{Jager:1998va}, which corresponds to $\lambda_5\to\infty$, with
$\lambda_5\sin^2\beta$ and $m_{12}^2\sin\beta$ fixed. In terms of the
parameterization of Eq.~(\ref{potentialprime}), used in
Ref.~\cite{Chivukula:1998zn}, this limit corresponds to $\lambda_5^\prime =
-\lambda_6^\prime\to\infty$, with $\lambda_5^\prime\sin^2\beta$ fixed.

\end{document}